**Remote Plasma Polymers of Iron (II) Phthalocyanine in Polyacrylonitrile-Derived Carbon Electrospun Fibers as Electrode for Supercapacitors.**

Jose M. Obrero [a*], Jorge PV Tafoya [b], Michael Thielke [b], G.P. Moreno-Martínez [a], Lidia Contreras-Bernal [a,c], Jose Ferreira de Sousa Jr [a], Juan Ramón Sánchez-Valencia [a], Angel Barranco [a], and Ana B. Jorge Sobrido [b*].

a) Nanotechnology on Surfaces and Plasma Laboratory, Materials Science Institute of Seville (CSIC-US), C/ Américo Vespucio 49, 41092, Seville, Spain.

b) Centre for Sustainable Engineering, School of Engineering and Materials Science, Faculty of Science and Engineering, Queen Mary University of London, Mile End Road, London, E1 4NS, UK

c) Química-Física. Department of Physical Chemistry, University of Seville, C/Professor García González n° 2, Seville 41012, Spain

E-mail: jmanuel.obrero@icmse.csic.es, a.sobrido@qmul.ac.uk

Keywords: supercapacitor, plasma processing, PAN fibers, electrospinning, plasma polymer, iron (II) phthalocyanine, remote plasma synthesis.

Remote plasma-assisted vapour deposition under nitrogen (RPAVD-$N_2$) is introduced as a single-step, solvent-free, room-temperature strategy to integrate iron(II) phthalocyanine (FePc) into carbon nanofiber (CNF) scaffolds for high-performance pseudocapacitive electrodes. In this process, CNFs are activated by low-energy $N_2$ remote plasma and subsequently exposed to sublimated FePc, which undergoes controlled plasma polymerisation to form conformal, nitrogen-rich FePc-derived coatings while preserving Fe–N coordination. By tuning the plasma power, the degree of crosslinking, defect generation and molecular fragmentation is precisely controlled. Structural and spectroscopic analyses reveal progressive incorporation of amine, nitrile and oxygenated functionalities while maintaining the Fe–N coordination environment, with 30 W power providing the optimal balance between structural integrity and defect density. Plasma processing enhances the capacitance by nearly one order of magnitude compared to sublimated FePc films, underscoring the critical role of plasma-induced molecular integration. The FePc30W@CNFs electrode delivers 80.9 $F \cdot g^{-1}$ at 0.25 $A \cdot g^{-1}$ (areal capacitance 0.92 $mF \cdot cm^{-2}$ at 2.9 $mA \cdot cm^{-2}$), achieves 7.42 $Wh \cdot kg^{-1}$ at 225 $W \cdot kg^{-1}$, and retains 86.5% of its initial capacitance after 6000 cycles. These results demonstrate that remote plasma polymerisation enables robust, high-rate and durable phthalocyanine-based electrodes, establishing RPAVD as a scalable platform for next-generation energy-storage materials.

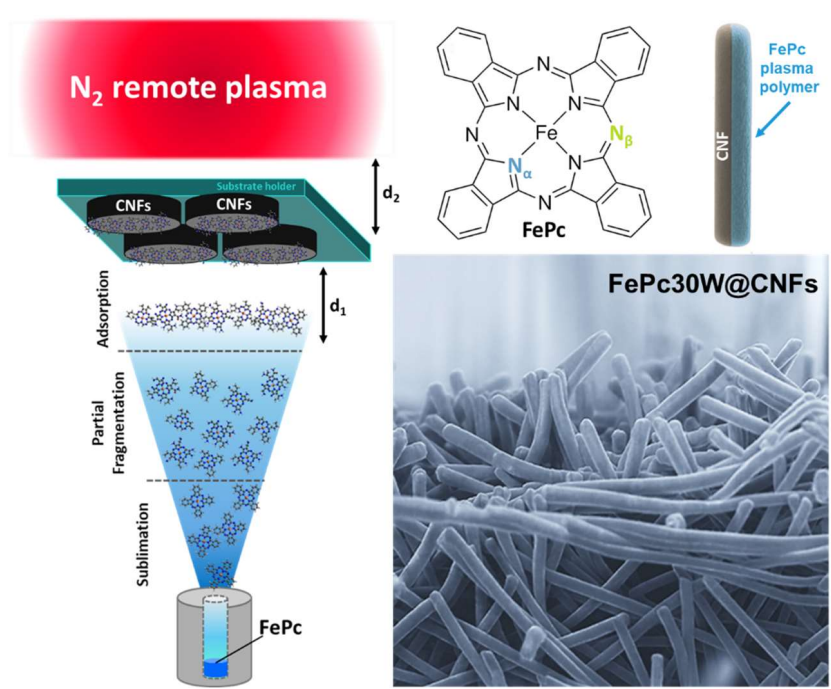

## 1. Introduction

The rapidly increasing demand for clean, efficient, and reliable energy storage technologies has become a critical challenge in the context of integrating renewable energy sources. In parallel, the rise of portable and wearable devices has highlighted the need for energy storage systems that can handle rapid cycling and high-power demands, beyond the capabilities of conventional batteries.[1] Supercapacitors have emerged as promising candidates to address this challenge, offering long cycle life, high power density, ultrafast charge-discharge capability, low maintenance costs, eco-friendliness, and exceptional cycling stability.[2] Nevertheless, their relatively low energy density compared to batteries continues to limit their broader implementation, driving intense research efforts toward advanced electrode materials and architectures. Based on their energy storage mechanisms, supercapacitors can be classified into electric double-layer capacitors (EDLCs) and pseudocapacitors.[3] The nature, structure, and accessibility of the electrode active materials fundamentally govern the electrochemical performance of supercapacitors. Carbon-based materials, including activated carbons, carbon nanotubes, and graphene derivatives, dominate both commercial and academic research owing to their high surface area, excellent electrical conductivity, and robust rate capability.[4] However, their charge-storage mechanism is primarily based on electric double-layer capacitance, which intrinsically restricts the achievable capacitance and energy density. In contrast, pseudocapacitors rely on fast and reversible surface or near-surface Faradaic reactions, enabling substantially higher capacitance. Representative pseudocapacitive materials include transition-metal oxides,[5] conductive polymers,[6] and redox-active molecular systems.[7] Despite their promise, these materials often suffer from limited electrical conductivity, inefficient utilisation of electroactive sites, structural degradation upon cycling, and difficulties in achieving homogeneous and stable coupling with conductive carbon matrices.[8,9] To overcome these limitations, composite electrode architectures combining conductive carbon frameworks with pseudocapacitive components have been widely explored, aiming to synergistically merge high electrical conductivity, large surface area, and enhanced redox activity. However, in most reported composites, the pseudocapacitive phase is physically mixed, weakly anchored, or heterogeneously distributed, which frequently results in poor electrical coupling, inefficient utilisation of molecular active sites, and limited long-term electrochemical stability. Therefore, developing composite electrodes that enable controlled, conformal integration of pseudocapacitive species within a conductive carbon network, independent of the specific material class, remains a critical challenge.[10]

Within this broader context, redox-active molecular systems integrated within conductive carbon frameworks represent an attractive subclass of pseudocapacitive materials.[7] In such hybrid architectures, the molecular units provide discrete redox-active centres, while the carbon scaffold ensures fast charge transport and mechanical stability. Among these molecular systems, metal phthalocyanines have gained increasing attention in recent years.[11–13] These complexes exhibit reversible metal-centred redox chemistry and an extended π-conjugated structure that facilitates efficient charge transfer and surface-confined Faradaic reactions. Electrochemical studies have shown that phthalocyanine-based electrodes can display characteristic pseudocapacitive behaviour, including quasi-rectangular voltammetric profiles, high rate capability, and good reversibility. To date, the incorporation of metal phthalocyanines into carbon-based electrodes has relied primarily on wet-chemical routes, physical mixing, or post-synthetic assembly, among others.[11,14–16] Although these approaches enable the introduction of molecular redox units into conductive matrices, they frequently result in heterogeneous molecular distribution, aggregation-induced loss of electroactive sites, and weak interfacial electronic coupling. Moreover, multistep fabrication procedures and harsh thermal or chemical treatments can compromise the intrinsic molecular structure of phthalocyanines, causing partial loss of metal-nitrogen coordination and reduced long-term electrochemical stability. As a result, current strategies offer limited control over molecular-level integration and interfacial chemistry, restricting the efficient utilisation of redox-active centres and hindering reproducible device performance. In parallel, plasma-based techniques have been increasingly explored for the fabrication and modification of supercapacitor electrode materials, owing to their ability to tailor surface chemistry, introduce heteroatom doping, engineer defects, and deposit functional coatings under solvent-free and low-temperature conditions.[17,18] Most reported plasma approaches have focused on the synthesis or activation of carbonaceous materials, including activated carbons, carbon nanotubes, graphene derivatives, and vertically aligned nanostructures.[19] Plasma processing has also been employed for the deposition of inorganic pseudocapacitive phases such as transition-metal oxides and nitrides.[20,21] However, despite the extensive use of plasma technologies in supercapacitor research, their application to the molecular-level integration of redox-active phthalocyanines into conductive carbon frameworks remains remarkably limited. In particular, plasma-enabled strategies capable of immobilising molecular pseudocapacitive units in a conformal and stable manner, while preserving their metal–nitrogen coordination environment and intrinsic redox activity, have been scarcely explored.

In this work, we introduce a novel plasma-enabled methodology based on remote plasma-assisted vapour deposition (RPAVD),[22–27] using nitrogen plasma to fabricate molecularly integrated pseudocapacitive electrodes based on iron(II) phthalocyanine (FePc) films. In this continuous single-reactor process, a carbon nanofiber (CNF) freestanding scaffold is initially exposed to a low-energy remote nitrogen plasma, resulting in surface activation and nitrogen functionalisation of the conductive carbon network. Subsequently, FePc molecules are sublimated into the downstream plasma region, where they undergo controlled plasma polymerisation to form nitrogen-rich plasma polymers conformally deposited onto the pre-functionalised carbon surface. This sequential dual functionality, achieved within a single uninterrupted plasma process, enables intimate interfacial coupling and enhanced electrical connectivity between the redox-active plasma polymer and the carbon scaffold. In this configuration, the carbon framework primarily contributes electric double-layer capacitance and electronic conductivity, whereas the FePc-derived plasma polymer introduces fast surface-confined Faradaic reactions, together resulting in a hybrid pseudocapacitive electrochemical capacitor behaviour. Importantly, the RPAVD is a solvent-less room-temperature process fully compatible with thermally or chemically sensitive substrates. By carefully tuning the plasma–molecule interaction, this strategy allows the immobilisation of FePc-derived plasma polymers with different degrees of polymerisation, preserving the Fe–N coordination environment and its intrinsic redox activity, while simultaneously promoting defect engineering and interfacial stability within the carbon framework. Compared to sublimated FePc, the plasma-polymerised electrodes exhibit a nearly tenfold increase in capacitance in some cases, demonstrating that plasma processing not only enables molecular immobilisation but also fundamentally enhances pseudocapacitive performance. Using FePc as a model system, we demonstrate that RPAVD converts fragile molecular redox units into robust pseudocapacitive composite electrodes exhibiting high capacitance, excellent rate capability, and remarkable cycling stability. Beyond this specific case, the present work establishes remote plasma polymerisation as a general and versatile platform for the single-process fabrication of molecularly engineered pseudocapacitive electrodes for next-generation energy storage devices.

## 2. Experimental.

### 2.1. Synthesis of CNFs from polyacrylonitrile by electrospinning.

9.5 wt.% polyacrylonitrile (PAN, $M_W$ 150 kDA, Sigma Aldrich) and 0.5 wt.% carbon black (C) (99.9%, 75 $m^2 \cdot g^{-1}$, Alfa Aesar) were dissolved in DMF (>99.9%, Fisher) overnight. Electrospinning was performed at 24 kV positive voltage and a flow rate of 0.5 $mL \cdot h^{-1}$ in a standard horizontal setup with a grounded collector at a tip-to collector distance of 15 cm. (**Figure 1a**). A total of 8 mL was spun to obtain the PAN/C fibres (CNFs), which were subsequently activated in a preheated box furnace for 2 h at 200 °C and finally carbonised in a tube furnace at 900 °C for 4 h in a $N_2$ atmosphere with a heating rate of 5 $°C \cdot min^{-1}$ to obtain the final carbon nanofibres (CNFs).

### 2.2 Deposition of FePc plasma polymers thin films by Remote Plasma Assisted Vapour Deposition (RPAVD).

Iron (II) phthalocyanine (FePc, with a dye content > 90%) was purchased by Sigma-Aldrich and utilised without further purification. CNF samples were positioned within an ECR-MW plasma reactor (**Figure 1b**) operating at 2.45 GHz, between the plasma discharge and the Knudsen sublimation cell, with the CNFs facing away from the glow discharge (the CNF-to-glow-discharge distance is 6 cm, and 12 cm in the most remote plasma-induced deposition). In addition to CNF electrodes, clean flat substrates of silicon (Si (100)) and fused silica were placed alongside the CNFs inside the reactor. These planar substrates served as reference supports to evaluate the morphology and optical properties of the FePc plasma polymer films under identical deposition conditions. Before initiating plasma processes, the reactor was evacuated to a base pressure of approximately $10^{-6}$ mbar. The CNFs were then subjected to an $N_2$ plasma pretreatment at 240 W for 30 minutes to generate surface defects and incorporate nitrogen-containing functionalities. The nitrogen flow was dosed with a calibrated mass flow controller set at 30 sccm. The gas pressure was controlled via an automated butterfly valve regulator, which controls the pumping flow of the system to maintain a pressure of $10^{-2}$ mbar. Immediately after pretreatment, FePc molecules were sublimated under $N_2$ plasma conditions with different plasma powers and geometrical configurations. During FePc sublimation, the working pressure was kept at $10^{-2}$ mbar. The experimental parameters selected for nitrogen plasma polymerisation, including plasma power, treatment distance from the sublimation cell, relative position to the glow discharge, and the sample labels assigned to each condition, are summarised in **Table 1**.

**Table 1.** Experimental conditions employed for the conformal deposition of FePc remote plasma polymers and sublimated films.

| Sample label | $N_2$ plasma power (W) | $d_{sample-sublim.}$* (cm) | $d_{sample-glow}$* (cm) |
|---|---|---|---|
| Sublimated FePc (FePcSubl) | 0 | 9 | - |
| FePc 30W-3cm | 30 | 3 | 12 |
| FePc 30W | 30 | 9 | 6 |
| FePc 60W | 60 | 9 | 6 |
| FePc 240W | 240 | 9 | 6 |

*$d_{sample-sublim.}$ and $d_{sample-glow}$ indicate the distance between the sample and the sublimation cell and the sample and the glow discharge region, respectively.

The deposition rate and resulting thickness of each FePc were monitored using a quartz crystal microbalance (QCM) placed near the sample holder. The temperature of the Knudsen cell was adjusted to maintain a constant growth rate of 0.5 Å·s$^{-1}$ (density of 0.5 g·cm$^{-3}$, and z-factor of 1.0 in the QCM electronics). The substrates remained at room temperature throughout deposition, as measured by an encapsulated thermocouple connected to the sample holder, which was unaffected by the plasma discharge. To ensure uniform plasma polymer coverage across the entire CNF network, including fibres located deeper within the porous architecture, the samples were flipped after the first deposition step. A second deposition was then performed on the previously shadowed reversed side under identical plasma conditions, resulting in homogeneous coating on both faces of the fibrous electrodes and throughout the three-dimensional network. In addition, the CNF substrates received the same $N_2$ plasma pretreatment on the hidden face before FePc deposition, as described above, so that both sides of the electrodes were functionalised under equivalent conditions.

*2.3 Characterisation methods.*

High-resolution field emission scanning electron microscopy (FESEM) images of the samples deposited on CNFs and silicon wafers were obtained in a Hitachi S4800 field emission microscope, working at 2 kV. Cross-sectional views of the flat substrates were prepared by cleaving the Si (100) wafers before imaging.

Optical transmittance properties of the samples deposited on fused silica substrates have been analyzed in the 200-150 nm wavelength range, recorded in a PerkinElmer Lambda 750 S UV–vis–NIR spectrophotometer. The spectral resolution was 2 nm.

Raman spectroscopy measurements were performed using a Renishaw inVia confocal Raman microscope equipped with a 20× objective lens and a 633 nm red laser, operated at 50% of its nominal power. Each spectrum was acquired with five accumulations to improve the signal-to-noise ratio. *In situ* Raman characterization of the electrospun mats was carried out using a dedicated electrochemical Raman cell (MTI Corporation, EQ-STC-Raman). In these experiments, the applied potential of the supercapacitor assembly was held until equilibrium was reached prior to spectral acquisition. The charging process was investigated by increasing the potential from an initial value of 0.00 V in equidistant steps of 0.1 V up to 0.75 V (last step of 0.05V). For the discharging process, the same procedure was followed by decreasing the applied potential from 0.75 V back to 0.00 V. The spectral resolution for this configuration was ~ 1.6 cm$^{-1}$. No polarisation was applied during the experiments.

XPS characterisations were performed using a Scientific Nexsa X-ray photoelectron spectrometer. The spectra were collected in the pass energy constant mode at 50 eV using a monochromated Al Kα X-ray source. C1s signal at 284.8 eV was used to calibrate the binding energies (BE). High-resolution spectra of individual elements (C1s, N1s, O1s, Fe2p, K2p) were collected with 50 accumulations to improve the signal-to-noise ratio. Peak assignments were performed according to values reported in the literature.[28,29]

Electrochemical measurements (cyclic voltammetry, CV; chronopotentiometry and electrochemical impedance spectroscopy, EIS) were performed using a supercapacitor assembly in a quick-assembly split cell (MTI Corporation, EQ-HSTC, 20 mm diameter). The full-cell assembly was connected to a Biologic SP300 potentiostat. For the physical assembly of the supercapacitor, two disks of 0.6 cm diameter were punched from the electrospun mats coated with plasma-polymerised FePc (FePc@CNFs). A third disk of equal diameter was cut from Whatman Grade GF/D Glass Microfiber Prefilters, used as the electrolyte reservoir. This separator disk was soaked overnight in 6 M KOH solution to ensure full impregnation with the electrolyte. Prior to assembly, the soaked separator was gently drained to remove excess solution. The final cell configuration consisted of a symmetric stack: FePc@CNFs / electrolyte-soaked separator / FePc@CNFs, compressed within the EQ-HSTC cell. The mass used for the calculation of the gravimetric capacitance corresponds to the total mass of the active electrode (CNFs + FePc samples) in the symmetric configuration. The mass of each electrode disk was measured using an analytical precision balance (±0.01 mg accuracy) prior to cell assembly. No subtraction of the pristine CNF mass was applied, as the performance of the full composite electrode was evaluated.

## 3. Results and discussion.

### 3.1 Characterisation of the FePc plasma polymers.

In the RPAVD technique, the solid precursor is sublimated into the downstream region of a microwave plasma discharge, where it interacts with reactive species at remote plasma conditions. As a result, the molecules are not fully fragmented, and a substantial proportion of intact or partially dissociated precursor units becomes incorporated into a growing polymeric film.[22–26,30,31] This process provides a versatile and solvent-free route to immobilise transition-metal macrocycles on conductive substrates at room temperature. Unlike our previously reported plasma polymerisation studies that used argon as the working gas,[22–26,30,31] the present work employs nitrogen (RPAVD-$N_2$). The use of $N_2$ plasma is particularly advantageous because it can promote the nitridation or nitrogen functionalisation of specific sites on the CNF substrates, improving the interaction between the plasma polymer and the carbon matrix. The degree of interaction between the plasma and the precursor, defined here as the polymerisation degree, can be tuned by adjusting the plasma power. In the present work, this parameter was varied between 30 and 240 W, together with the control of the distance between the sample and glow discharge (denoted as X in **Figure 1b**, see also **Table 1**), allowing the systematic study of its effect on the structural, chemical, and electrochemical properties of the resulting FePc plasma polymers.

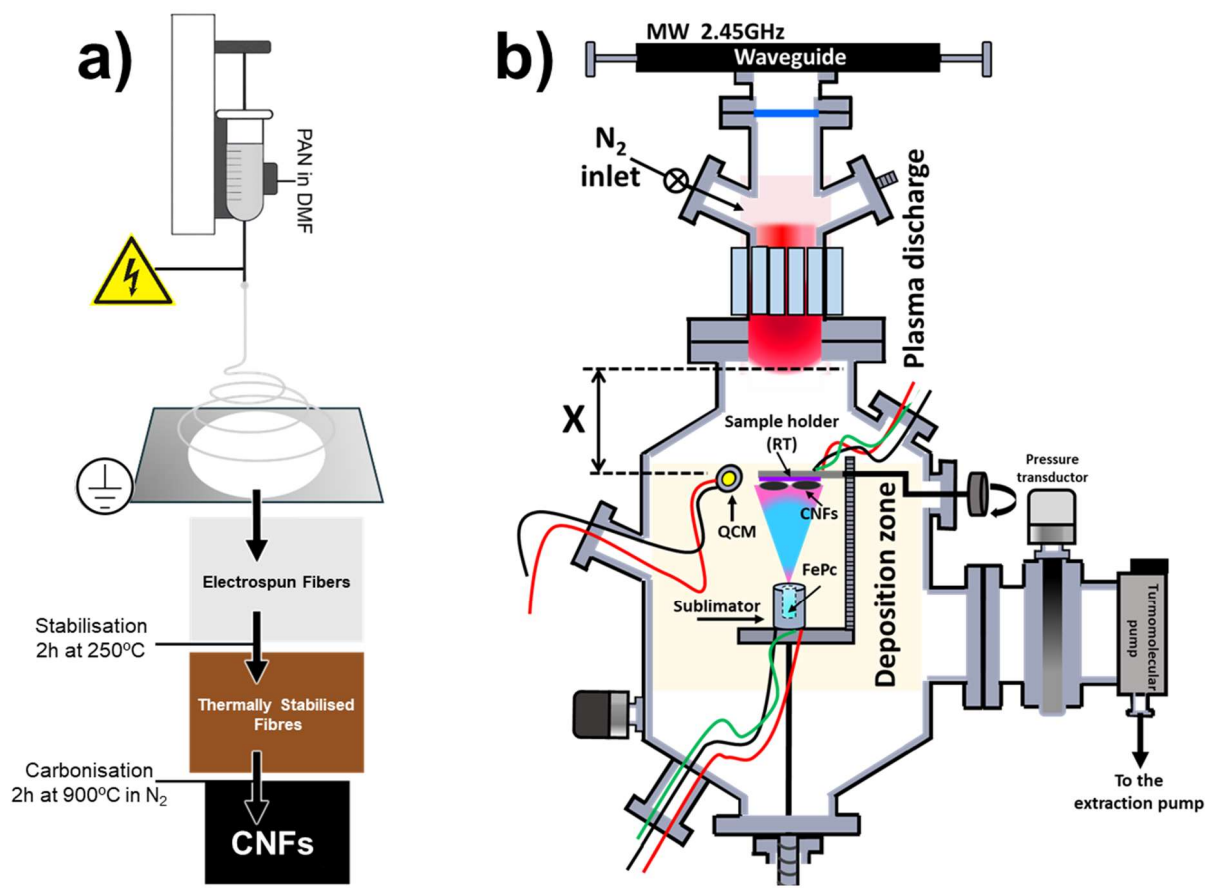

**Figure 1.** Schematic representation of the fabrication process for FePc plasma polymers@CNFs. (a) Schematic of the conventional electrospinning process for preparing CNFs. A droplet of PAN/C dissolved in DMF is subjected to a high electric field, producing a charged jet that is accelerated toward a grounded collector. The resulting fibre mat is removed, thermally stabilised, and carbonised to obtain CNFs. (b) Schematic of the RPAVD reactor used for plasma-assisted polymerisation of FePc.

The SEM images of the RPAVD-$N_2$ samples on CNFs obtained at different polymerisation degrees (including the sublimated one) are shown in **Figure 2**. The as-prepared CNFs **(Figure 2a)** exhibit a smooth and clean surface with almost no visible roughness. The slight surface irregularities observed are mainly attributed to the preliminary $N_2$ plasma exposure, a routine pretreatment that introduces surface defects and nitrogen-containing functional groups, thereby enhancing surface reactivity and adhesion of the subsequent FePc coating. A comparison of the CNFs before and after the $N_2$ plasma pretreatment is presented in **Figure S1**. The direct sublimation of FePc produces a polycrystalline conformation (**Figure 2b**) with randomly oriented grains formed by the metallic complex molecules. The RPAVD of FePc onto the CNFs generates smooth samples with almost inappreciable features at the nanoscale (**Figures 2c–e**).

In all these cases, the $N_2$ plasma promotes the formation of a uniform shell surrounding each CNF, showing excellent homogeneity in thickness from the top to the bottom of the fibres. At low plasma power (30 W), the FePc plasma polymer displays a rougher and more granular morphology, particularly evident in the film deposited on the Si substrate (**Figure 2f**). Although the coating on the flat Si substrate reaches an average thickness of approximately 210 nm, the nominal thickness of the FePc plasma polymer layer on CNFs varies depending on the specific processing conditions. A statistical distribution of CNFs diameters at different RPAVD power is shown in **Figure S2**. The granular morphology observed at 30 W indicates a limited interaction between the FePc precursor and the plasma, enabling the deposition of only partially polymerised species and precursor crystallites. Increasing the plasma power to 60 W and 240 W progressively smooths and compacts the surface, consistent with enhanced molecular fragmentation and crosslinking. The stronger interaction with the $N_2$ plasma promotes greater precursor activation and film densification, yielding a more homogeneous and continuous layer. Overall, these results show that higher plasma power increases the degree of FePc polymerisation, producing denser but less molecularly defined coatings, as observed in the film on Si(100) substrate shown in **Figure S3**.

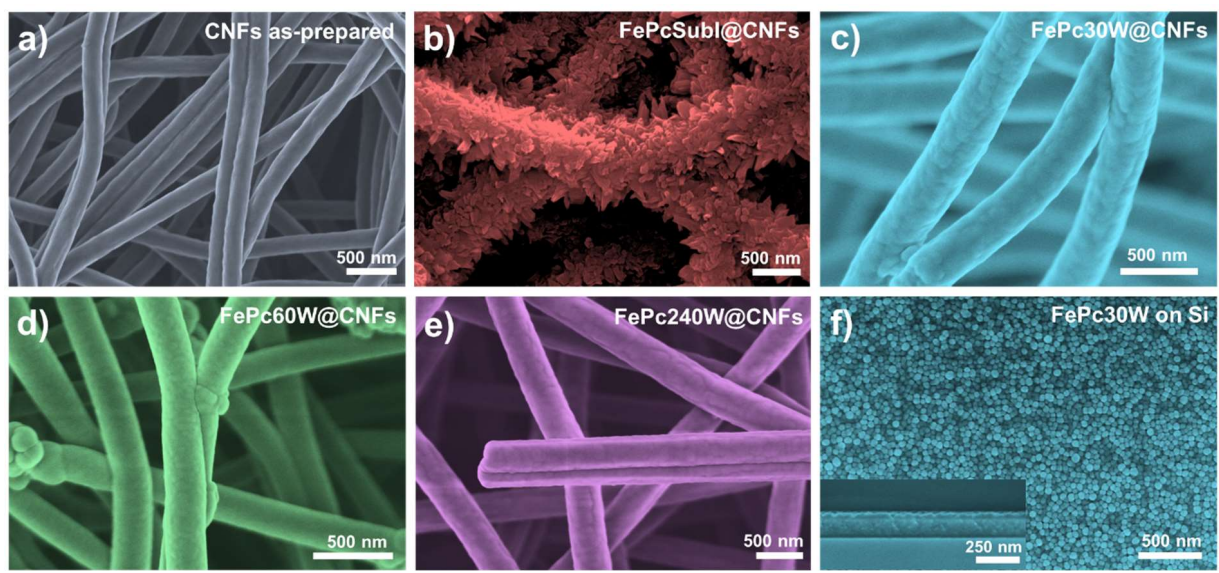

**Figure 2.** SEM micrographs of the CNF-based systems obtained by deposition FePc by RPAVD-$N_2$ at different polymerisation degrees. (a) CNFs; (b) CNFs coated with FePc by sublimation; (c) RPAVD at 30 W; (d) at 60 W; (e) at 240 W; and (f) top-view of the FePc by RPAVD at 30 W deposited on a flat (silicon) substrate. The inset in (f) shows the corresponding cross-sectional view of the same film.

The optical response and molecular integrity of the FePc plasma polymers were analysed by UV-vis-NIR spectroscopy (**Figure 3a**). The sublimated FePc film shows the characteristic absorption bands of iron(II) phthalocyanine, with a Soret band at 621 nm (Q-band region) accompanied by a small shoulder at 564 nm, and a B band at 315 nm, confirming the preservation of the molecular structure.[32,33] Upon plasma polymerisation via RPAVD-$N_2$, these spectral features progressively diminish in intensity and undergo bathochromic shifts as the plasma power increases. In the FePc 30W-3cm sample, both the Soret and B bands remain nearly unchanged compared to the sublimated film, suggesting minimal disruption of the FePc macrocycle. However, at higher plasma powers, notably in the FePc 240W sample, the Soret band shifts markedly to 663 nm, accompanied by a significant reduction in intensity and complete loss of the B band. This spectral evolution reflects a gradual loss of molecular order and $\pi$-conjugation, as well as the possible partial oxidation of the central iron atom. These changes indicate a transition from a well-defined molecular material (intact FePc) to a crosslinked and partially delocalized plasma polymer network. In contrast to the sublimated FePc film, which undergoes progressive degradation upon air exposure, all plasma polymer films retain their optical integrity over time. This enhanced environmental stability is an intrinsic feature of plasma polymers and makes them particularly suitable as protective and functional coatings.[30,34] Evidence of this effect is presented in **Figure S4a**, where the sublimated FePc film exhibits the appearance of a shoulder at ≈710 nm after 1300 days of air exposure, indicative of oxidation and molecular degradation.[35,36] Conversely, the RPAVD-$N_2$ 30W plasma polymer **(Figure S4b)** shows no detectable spectral changes after the same period, confirming its remarkable long-term stability.

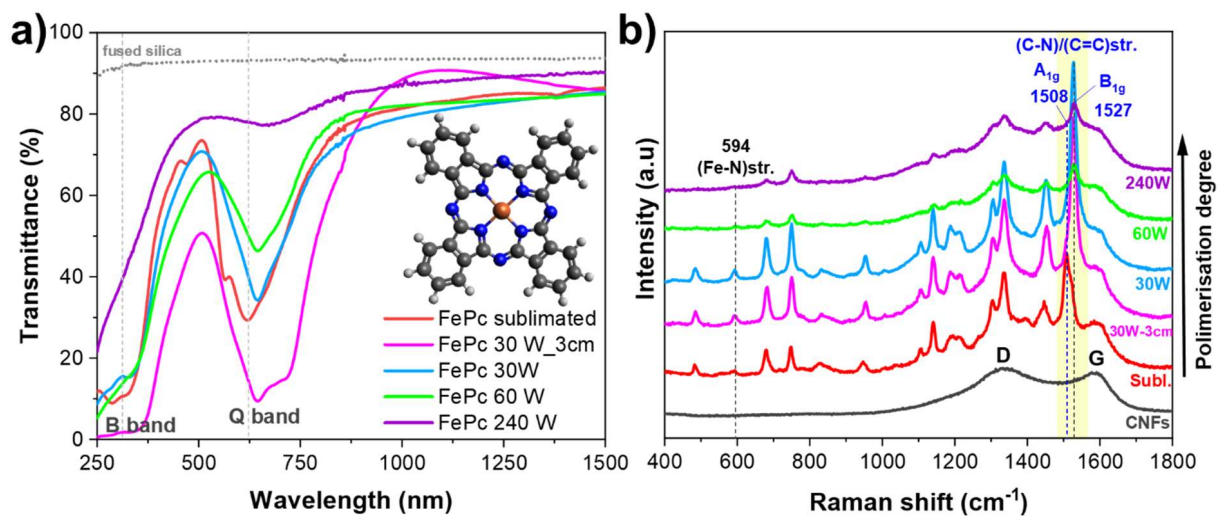

**Figure 3.** (a) UV–Vis–NIR transmittance spectra of FePc thin films deposited on fused silica, comparing sublimated samples with those plasma-polymerised via RPAVD-$N_2$ at different degrees of polymerisation. (b) Raman spectra of FePc@CNFs samples at varying polymerisation degrees, including the reference spectrum of bare CNFs.

Raman spectra of the thin films of the FePc sublimated and plasma polymer films deposited on CNFs (FePc@CNFs) are presented in **Figure 3b**. The characteristic D and G bands of carbon-based materials are clearly observed at approximately 1336 cm$^{-1}$ and 1586 cm$^{-1}$, respectively. The intensity ratio $I_D/I_G$ is approximately 1.10, indicating a high degree of structural disorder within the fibres. This value exceeds the typical range reported for similar systems, where $I_D/I_G$ ratios generally fall between 0.75 and 0.81,[37] which suggests that the increased disorder is attributed to the mild $N_2$ plasma pretreatment applied to the fibres. In addition, a significant broadening of the D band is observed, suggesting a wide distribution of graphene-like domain sizes and elevated structural heterogeneity.[38] This phenomenon supports the hypothesis that plasma exposure induces controlled fragmentation of the graphitic network, resulting in a more reactive surface architecture. Collectively, these effects are expected to enhance ion adsorption and, consequently, improve the specific capacitance of the material.

The FePc$_{sublimated}$@CNFs sample (red spectrum) shows the fingerprint Raman bands of iron(II) phthalocyanines.[39,40] Among its main signals, the band at ~594 cm$^{-1}$ is associated with Fe–N stretching, confirming the coordination of the metal centre within the macrocycle. Other relevant peaks are found at 482, 680, 750, 832, 950, 1140, 1187, 1306, 1336, and 1452 cm$^{-1}$, corresponding to vibrational modes of the macrocyclic ring, mainly C–N and C=C stretching and in-plane deformation modes. In the high-wavenumber region, two prominent bands appear at ~1508 and ~1527 cm$^{-1}$. According to the literature, these features correspond to symmetry-allowed skeletal vibrations of the phthalocyanine macrocycle, namely the totally symmetric $A_{1g}$ mode (~1508 $cm^{-1}$) and the $B_{1g}$ mode (~1527 $cm^{-1}$), both arising from collective C–N/C=C stretching of the isoindole and benzene rings rather than from distinct local functional groups.[39,40] In well-ordered crystalline FePc films, the higher-frequency $B_{1g}$ mode around 1530-1540 $cm^{-1}$ is typically the most intense and is often used as a marker of the metallophthalocyanine framework.[41,42] In contrast, the sublimated FePc film deposited on CNFs shows a dominant contribution at ~1508 $cm^{-1}$ ($I_{1508}/I_{1527} \approx 3.0$), indicating a relative enhancement of the totally symmetric $A_{1g}$ mode. This behaviour can be attributed to the disordered molecular arrangement and substrate interaction in the sublimated layer, which relax Raman selection rules and modify the relative intensities of symmetry-dependent vibrational modes. After mild $N_2$ plasma treatment (30 W–3 cm and 30 W), a clear redistribution of intensity is observed: the ~1527 $cm^{-1}$ $B_{1g}$ band becomes dominant and the $I_{1508}/I_{1527}$ ratio drops to ~0.06. This change occurs due to a plasma-induced reduction of molecular symmetry and the initiation of crosslinking within the phthalocyanine network. At higher plasma power (60 W and 240 W), the ~1527 $cm^{-1}$ band remains predominant but becomes broader and less defined, indicating increased structural disorder and partial amorphisation due to excessive crosslinking. The Fe–N stretching band at ~594 $cm^{-1}$ remains visible in all samples, confirming that the Fe centre stays coordinated and that no demetalation occurs during plasma processing. Overall, mild plasma polymerisation induces a controlled reorganisation of the FePc macrocycles, whereas stronger plasma conditions lead to a loss of long-range order without disrupting the metal–nitrogen coordination, anticipating the electrochemical behaviour discussed in the following section.

The surface chemical composition and stoichiometry of the FePc sublimated and plasma polymer films deposited on CNFs, as well as of the pristine CNF substrates, were analysed by X-ray photoelectron spectroscopy (XPS). The corresponding survey spectra and the atomic percentages of each element are shown in **Figure 4a and f**, respectively. As the degree of plasma polymerisation increases, the films exposed to the $N_2$ plasma show a progressive increase in the relative nitrogen content compared to the sublimated FePc film. Starting from an $(N/C)_{sublimated}$ ratio of 0.15, close to that expected from the empirical formula of FePc ($C_{32}H_{16}N_8Fe$), the N/C ratios of the plasma polymer films increase to $(N/C)_{30W-3cm} = 0.23$; $(N/C)_{30W} = 0.28$; $(N/C)_{60W} = 0.33$ and $(N/C)_{240W} = 0.50$. This trend confirms the effective incorporation of nitrogen into the polymeric matrix as the plasma-precursor interaction becomes stronger, while the overall carbon content gradually decreases. In all samples, a certain amount of oxygen is also detected. In the case of the sublimated FePc film, the oxygen traces are attributed to mild surface oxidation upon air exposure after deposition and water adsorbed on the surface. For the plasma polymer films, however, the higher oxygen content is a characteristic feature of any plasma-assisted process, such as RPAVD, which can incorporate residual oxygen from the reactor atmosphere during deposition and from post-deposition reactions of dangling bonds and trapped radical species with ambient air.[23,30] On the other hand, the amount of atomic Fe remains nearly constant for all samples, close to 1 at.%, indicating that the plasma treatment does not lead to significant loss or sputtering of iron.

As shown in **Figure S5**, $N_2$ plasma pretreatment significantly modifies the surface chemistry of the CNFs, increasing the relative amount of pyrrolic-N species. This nitrogen enrichment indicates a soft-etching process that generates reactive carbon sites subsequently functionalized by $N_2$ species.[43] Such changes are expected to influence the coordination behaviour of FePc upon deposition.

To evaluate the chemical evolution of FePc under different RPAVD polymerization degrees, high-resolution XPS spectra in the Fe2p, O1s, N1s and C1s regions were analysed (**Figure 4**). **Figure 4b** shows the XPS spectra corresponding to the Fe2p region, whose deconvolution reveals two main components across all samples: a peak at 709.4 eV attributed to $Fe^{2+}$ coordinated to nitrogen atoms of the phthalocyanine macrocycle (Fe–N), and a second peak at 712.5 eV associated with $Fe^{3+}$ species, resulting from partial oxidation of $Fe^{2+}$ to $Fe^{3+}$ (Fe–O)[44–46], as well as their spin-orbit counterparts separated by 13.1 eV. The Fe–N/Fe–O ratios (**Figure 4g**) reveal that the 30 W sample preserves the $Fe^{2+}$ coordination within the phthalocyanine macrocycle most effectively, whereas higher plasma powers progressively increase Fe-O formation, indicating stronger oxidation. This trend is corroborated by the O1s spectra (**Figure 4c**), which show contributions from $O_{Fe}$, $O_V$, $O_{OH}$, $O_C$ and $O_{H2O}$.[25,47] The $O_{Fe}$ component at 529.6 eV increases with power, consistent with enhanced Fe oxidation. The peak at 530.8 eV ($O_V$) corresponds to oxygen vacancies, which create defect sites that may favour charge transport but also indicate structural disorder when excessive. This behaviour is consistent with reports showing that oxygen vacancies improve conductivity, ion transport and the density of electrochemically active sites.[48,49] In contrast, the $O_{OH}$ peak at 531.8 eV rises sharply at high powers, reflecting excessive surface hydroxylation and partial passivation. The $O_C$ component at 532.8 eV corresponds to oxygen bound to carbon atoms in oxidized organic groups (C=O, -O-C=O), reflecting the partial oxidation of the phthalocyanine polymer backbone. Finally, the peak at 534.1 eV ($O_{H2O}$) is attributed to molecular water adsorbed on the surface. The relative ratios $O_{Fe}/O_T$, $O_V/O_T$ and $O_{OH}/O_T$ (**Figure 4h**) confirm that the 30 W condition provides the best compromise between controlled oxidation, defect creation and surface stability. The N1s spectra (**Figure 4d**) of sublimated FePc exhibit the expected pyridinic ($N_\beta$) and pyrrolic ($N_\alpha$) components, whose nearly identical integrated areas reflect the characteristic 1:1 abundance of these nitrogen environments in the phthalocyanine macrocycle, together with the typical $\pi$–$\pi$ satellite.[23,50] Plasma interaction induces small shifts while preserving the $N_\alpha/N_\beta$ ratio, indicating that the phthalocyanine macrocyclic framework is largely preserved**.** Two new components at 397.9 and 400.3 eV grow with increasing plasma power, arising from primary amines, nitriles, tertiary amines, and oxidised nitrogen species generated during plasma-driven fragmentation and reorganisation.[25,51,52] However, the strong overlap of nitrogen species in this region prevents an unambiguous assignment of the 400.3 eV peak, which likely reflects a mixture of amine, nitrile and oxidised nitrogen groups formed partly through secondary reactions with residual oxygen. This evolution reflects the enhanced incorporation of amine-, nitrile- and oxygen-containing nitrogen species, consistent with plasma-induced fragmentation and reorganisation of the FePc framework. These structural transformations are expected to influence the capacitive behaviour of the films, as discussed in the following section. Similarly, the C1s spectra (**Figure 4e**) evolve from the characteristic FePc structure to increasingly nitrogen- and oxygen-functionalised carbon environments. The C=N-related peak shifts by +0.6 eV due to reduced electron density, and two new peaks appear at 286.1 and 289.0 eV, corresponding to C–N groups and to oxygen-rich carbon species, respectively,[53,54] confirming the incorporation of amine, nitrile and oxidised groups. Since the binding energies of C–N and C=N overlap in the 285.9–287.0 eV range, their distinction in the C1s spectra is not straightforward; therefore, the assignments were confirmed through the N1s region, which more clearly reveals the formation of amine and nitrile groups under plasma exposure. Overall, XPS results demonstrate that increasing the plasma power during RPAVD-$N_2$ progressively introduces amine, nitrile and oxygenated functionalities, increases Fe oxidation and generates oxygen-related defects. Among the studied conditions, the 30 W sample displays the most favourable balance between structural preservation, controlled oxidation and defect formation, features that underpin its superior capacitive behaviour.

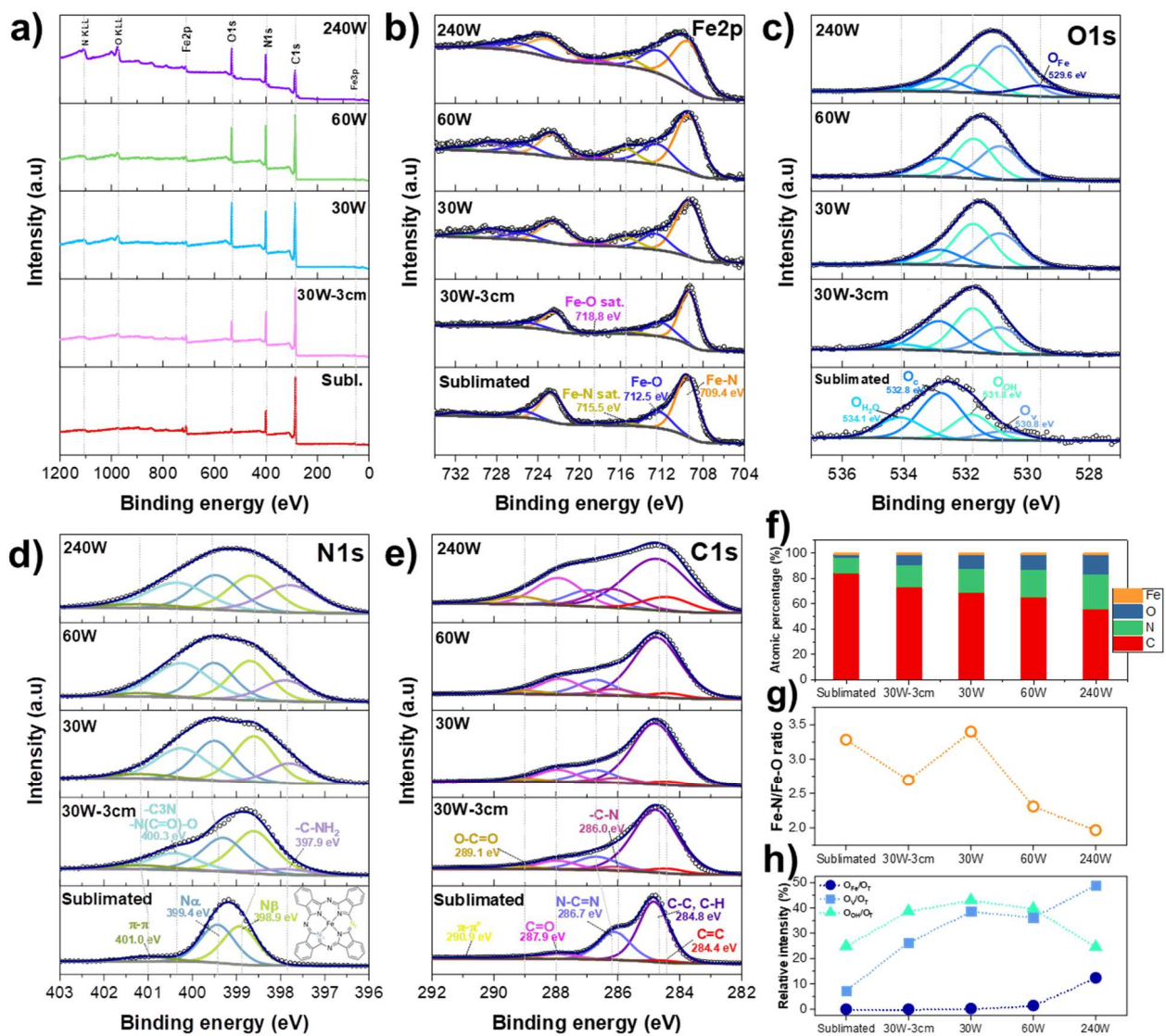

**Figure 4.** (a) XPS survey spectra of the sublimated FePc and the plasma-polymerised samples at different degrees of polymerization via RPAVD. Detailed high-resolution spectra of (b) Fe2p; (c) O1s; (d) N1s and (e) C1s regions. (f) Surface atomic percentages of the films determined by XPS. (g) Fe–N/Fe–O ratio as a function of polymerization degree extracted from the Fe2p region. (h) Relative contributions of $O_{Fe}$, $O_v$ and $O_{OH}$ species with respect to the total oxygen content ($O_T$), obtained from the deconvolution of the O1s spectra.

The electrochemical performance of each FePc-coated CNF (FePc@CNFs) film was evaluated using a two-electrode system in 6 M KOH aqueous solution. To demonstrate the capacitive behaviour of each sample, cyclic voltammetry (CV) and galvanostatic charge-discharge (GCD) measurements were compared. The electrochemical performance of the bare CNFs pre-treated with $N_2$ plasma is shown in **Figure S6**. The CNFs display a nearly rectangular CV curve, indicating that energy storage occurs predominantly through electric double-layer capacitance.[37] **Figure 5a** shows the CV response (the final voltammogram out of ten recorded) for each sample, measured at a scan rate of $10 \text{ mV·s}^{-1}$. As observed, the voltammogram of the FePcsublimated@CNFs sample (red trace) does not exhibit capacitive behaviour, displaying well-defined oxidation/reduction peaks within the studied potential window (0.00-0.75 V), characteristic of faradaic battery-type processes. These results are consistent with the XPS data, which indicate redox deactivation of FePc due to Fe–CNF coordination. The complete electrochemical performance of the sublimated FePc is presented in **Figure S7**. Following RPAVD polymerisation, this behaviour is no longer observed within the same potential range, and the resulting voltammograms become nearly rectangular, with the current remaining almost constant throughout the sweep and no discernible peaks. This response is consistent with pseudocapacitive behaviour, where charge storage arises from fast, reversible surface redox reactions rather than purely electrostatic double-layer formation. In metal phthalocyanines, this pseudocapacitive behaviour originates primarily from electron gain and loss at the central metal atom, and partly from proton absorption and desorption on pyridinic ($N_\beta$) or pyrrolic ($N_\alpha$) nitrogen sites.[55] In the case of FePc, the high pseudocapacitance can be attributed to the pronounced redox activity of the central Fe, as illustrated in Equation (1):

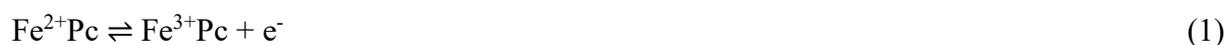

$$Fe^{2+}Pc \rightleftharpoons Fe^{3+}Pc + e^- \qquad (1)$$

This intrinsic redox activity explains the reproducible pseudocapacitive performance observed across different FePc electrodes and supports the enhanced charge storage capability of the optimised plasma-polymerised samples.

Although this behaviour is observed in most polymerised samples, not all exhibit the same electrochemical performance. In particular, the FePc30W@CNFs (blue) and FePc30W-3cm@CNFs (pink) samples, synthesised at the same plasma power but at different glow-discharge distances, exhibit the most pronounced pseudocapacitive behaviour, with nearly rectangular voltammograms enclosing the largest areas. In contrast, increasing the plasma power leads to a gradual loss of charge-storage capability: the FePc60W@CNFs sample (green) already shows a reduced enclosed area, and this effect becomes even more pronounced in the FePc240W@CNFs sample (purple), whose voltammogram becomes narrower and more oval-shaped. This progressive decline in pseudocapacitance correlates with the XPS results, which reveal increasing $Fe^{3+}$ content and diminished preservation of the Fe–FePc coordination environment.

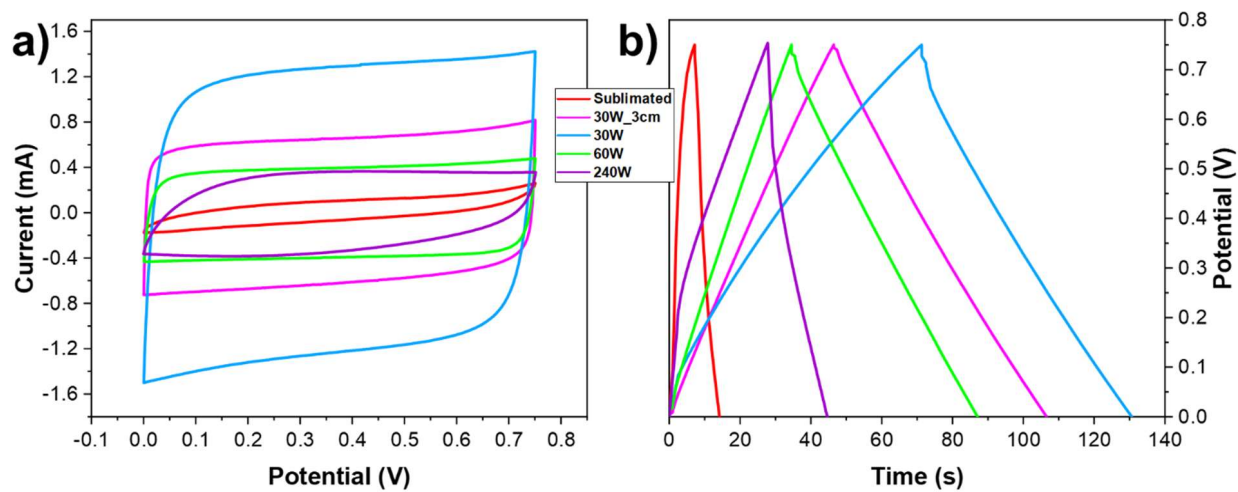

**Figure 5.** Electrochemical performance of FePc@CNFs films. (a) Cyclic voltammetry curves recorded at a scan rate of $10\,mV \cdot s^{-1}$. (b) Galvanostatic charge-discharge profiles measured at a current density of $0.25\,A \cdot g^{-1}$.

Complementary insight into the capacitive behaviour was obtained from the GCD curves at $0.25 A \cdot g^{-1}$ shown in **Figure 5b**. The FePc$_{sublimated}$@CNFs sample shows only a narrow, steep discharge profile, consistent with its poor capacitive response in CV. In contrast, the polymerised FePc@CNFs films exhibit nearly linear and symmetric triangular curves, confirming their pseudocapacitive behaviour and highly reversible redox processes. The longest discharge times are observed for FePc30W-3cm@CNFs and FePc30W@CNFs, in agreement with CV results and indicative of superior charge-storage capability. Increasing plasma power leads to shorter discharge times: the effect is already evident at 60 W and becomes more pronounced in FePc240W@CNFs, whose curve also deviates from linearity due to higher internal resistance. Overall, these results show that moderate plasma polymerisation preserves the electroactive FePc structure while enhancing conductivity through the CNF network, whereas excessive power results in molecular degradation and performance loss.

Overall, the comparative electrochemical analysis indicates that FePc30W@CNFs is the most promising condition. Its combination of highly pseudocapacitive signatures in both CV and GCD, together with the favourable structural features revealed by XPS, indicates that moderate plasma polymerisation at 30 W preserves the electroactive FePc environment while ensuring good electrical connectivity within the CNF network. Consequently, the following section focuses exclusively on FePc30W@CNFs, including its specific capacitance, cycling stability, and in-situ Raman response under applied potential.

## 3.2 Electrochemical performance of FePc30W@CNFs samples.

**Figure 6** summarises the electrochemical performance of the FePc30W@CNFs sample. The CV curves recorded at scan rates between 1 and $20\,mV \cdot s^{-1}$ (**Figure 6a**) retain a nearly rectangular shape, indicative of efficient pseudocapacitive behaviour. At higher scan rates ($50\,mV \cdot s^{-1}$), however, the profiles progressively become more oval, reflecting increasing ion-diffusion limitations and a corresponding decrease in charge-storage efficiency. The total capacitance stored in an electrode arises mainly from two contributions: (1) the surface-controlled capacitive phenomenon, which includes both the faradaic pseudocapacitance caused by fast surface redox reactions and the non-faradaic double-layer charging (capacitive contribution), and (2) the diffusion-controlled capacitive process, in which ion diffusion into the bulk of the electrode plays a dominant role (diffusive contribution). To quantitatively analyse the relative contributions of these capacitive elements, a well-established deconvolution method was applied.[56–58] The current response of a pseudocapacitive electrode at a given potential can be expressed as the sum of a surface-controlled term, proportional to the scan rate $v$, and a diffusion-controlled term, proportional to $v^{1/2}$:

$$i(v) = k_1 \cdot v + k_2 \cdot v^{1/2} \tag{2}$$

Here, the ($k_1 \cdot v$) component includes both double-layer charging and fast surface-controlled pseudocapacitance, while the ($k_2 \cdot v^{1/2}$) term reflects the slower faradaic processes limited by ion diffusion into the electrode bulk. The results obtained from the least-squares fitting of the pseudocapacitive model at different scan rates are shown in **Figure 6b**. The current values were extracted from the CV data at 0.4 V in the forward scan, a potential located within the pseudocapacitive region and free from redox peak interference. The excellent correlation between the model and the experimental data (**Figure 6c**), with $R^2 = 0.9948$, confirms the reliability of the deconvolution approach and validates the separation of capacitive and diffusive contributions. Additional fits at 0 V, 0.4 V (reverse), and 0.75 V (**Figure S8**) show the same behaviour, with equally high determination coefficients (see Table S1), demonstrating that the model (Equation 2) is robust across the entire potential window. At low scan rates, the current is dominated by diffusion-controlled processes, as ions have sufficient time to penetrate the electrode bulk and participate in faradaic reactions. At higher scan rates, the capacitive component becomes dominant because the rapid potential sweep favours surface-controlled charge-storage mechanisms over ion diffusion. This transition is reflected in **Figure 6b–d**, where the capacitive fraction reaches 62 % at 10 mV·s⁻¹, highlighting the major contribution of fast and reversible surface redox reactions in FePc30W electrodes. **Figure 6d** further illustrates the contribution of surface capacitance to the total capacitance in the CV profile recorded at 10 mV·s⁻¹, where the shaded region represents the capacitive fraction estimated from Equation (2).

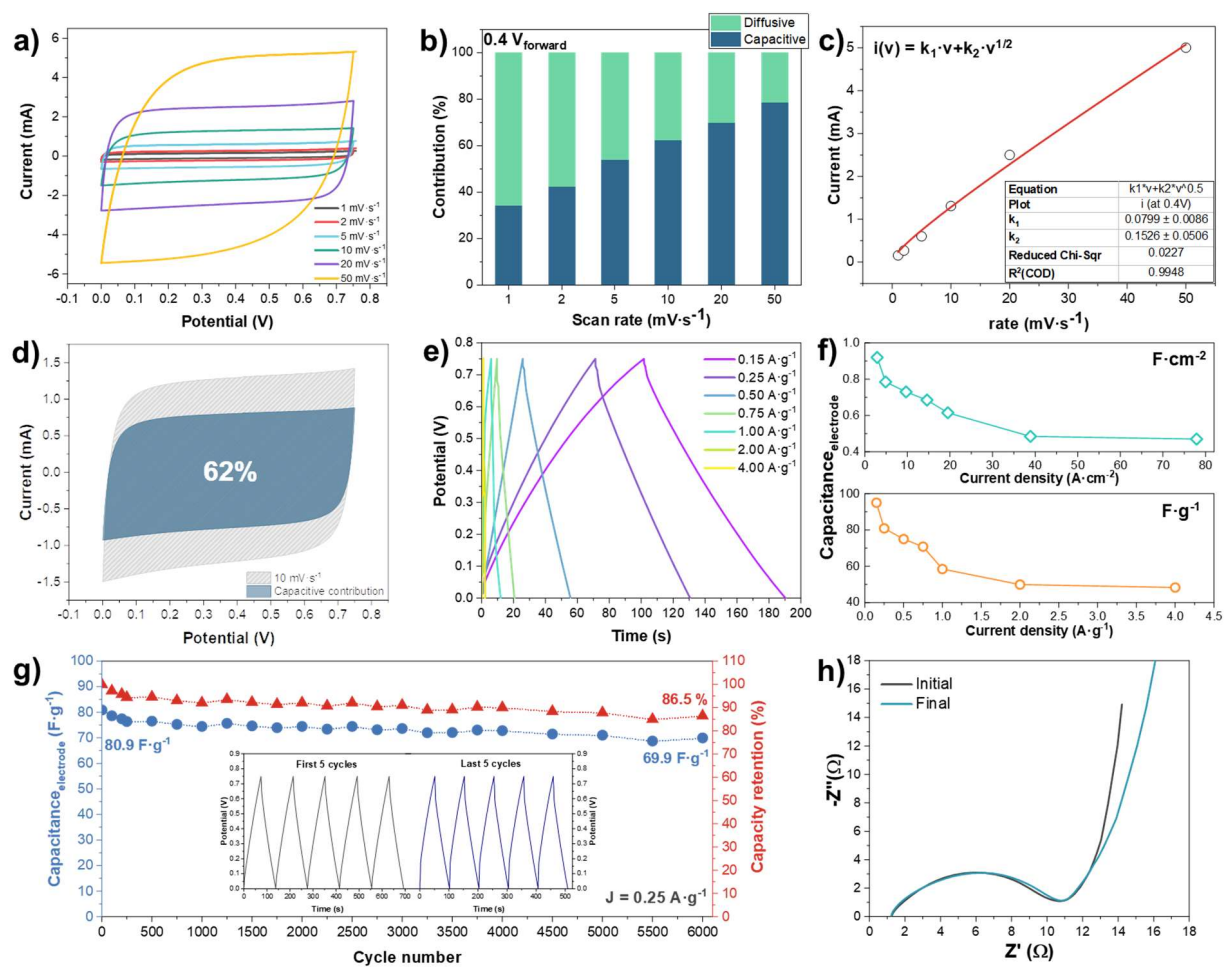

**Figure 6.** Electrochemical performance of FePc30W@CNFs sample. (a) CV curves recorded at scan rates ranging from 1 to 50 mV·s⁻¹. (b) Capacitive and diffusive contributions at different scan rates, extracted using the pseudocapacitive current model. (c) Least-squares fitting of the model i(v)=k₁·v+k₂·v^(1/2) applied to current data extracted from CV at 0.4 V in the forward scan. The inset displays the fitted k₁ and k₂ values along with the corresponding R². (d) CV curve at 10 mV·s⁻¹ highlighting the capacitive contribution. The shaded area is calculated according to Equation (2). (e) GCD curves recorded at various current densities. (f) Specific capacitance (C_electrode) expressed both gravimetrical (F·g⁻¹ vs A·g⁻¹) and areal (F·cm⁻² vs mA·cm⁻²) as a function of current density. (g) Cycling stability and capacitance retention at 0.25 A·g⁻¹. The inset shows the first and last five cycles during the stability test. (h) Nyquist plots recorded before and after cycling.

The electrochemical performance was further examined through GCD measurements. **Figure 6e** shows the GCD curves of the FePc30W@CNFs electrode at current densities from 0.15 to 4.0 A·g$^{-1}$ within the selected potential window. The profiles display an almost symmetric triangular shape, confirming the pseudocapacitive behaviour associated with fast surface-controlled faradaic reactions. The charging segments remain nearly linear, indicating rapid charge accumulation with minimal kinetic limitations, whereas the slight curvature during discharge reveals the participation of slower diffusion-controlled processes. This asymmetry highlights the coexistence of surface pseudocapacitance and diffusion-limited ion release, fully consistent with the CV findings. The specific capacitance was then calculated from the discharge time, where the gravimetric capacitance of the symmetrical cell ($C_{cell}$) and the single electrode ($C_{electrode}$) are defined as:

$$C_{symCell\,(gravimetric)} = \frac{i \cdot \Delta t_{discharge}}{M_{total} \cdot \Delta V} \tag{3}$$

$$C_{electrode\,(gravimetric)} = 4 \cdot C_{symCell\,(gravimetric)} \tag{4}$$

where $i$ is the discharging current, $\Delta t_{discharge}$ is discharging time, $M_{total}$ represents the total mass of the two active electrode materials in the symmetric cell and $\Delta V$ is the potential window. In addition, the areal capacitance of the electrode was calculated using the following expression:

$$C_{electrode\,(areal)} = \frac{2 \cdot i \cdot \Delta t_{discharge}}{A_{electrode} \cdot \Delta V} \tag{5}$$

where $A_{electrode}$ is the geometric area of one electrode. The variation of specific capacitance with current density is presented in **Figure 6f**. The FePc30W@CNFs electrode delivers 95.0, 80.9, 75.0, 70.8, 58.5, 49.9, and 48.3 F·g$^{-1}$ at current densities ranging from 0.15 to 4.00 A·g$^{-1}$, corresponding to areal capacitances of 0.92, 0.79, 0.73, 0.69, 0.62, 0.49, and 0.47 F·cm$^{-2}$ at 2.9-77.9 mA·cm$^{-2}$. These values far exceed those of the FePc$_{sublimated}$@CNFs electrode (8.5 F·g$^{-1}$ and 0.10 F·cm$^{-2}$ at 0.25 A·g$^{-1}$ and 3.1 mA·cm$^{-2}$), representing an enhancement factor of ~9.5 and confirming the superior charge-storage capability of the plasma-polymerised films. Comparing both gravimetric and areal capacitances shows that the plasma-polymerised FePc coating clearly outperforms the sublimated FePc layer. Its higher capacitance and better rate capability arise from the improved conductivity and structural robustness of the plasma-polymerised network on the CNFs. Cycling stability is also a key parameter for evaluating the practical applicability of supercapacitors,[59] and was assessed through repeated GCD measurements (**Figure 6g**). After 6000 cycles at 0.25 A·g⁻¹, the FePc30W@CNFs electrode retains 86.5% of its initial capacitance, demonstrating excellent long-term durability. Comparable retentions of 86.1% and 87.5% at 0.50 and 1.00 A·g⁻¹, respectively (**Figure S9**), further confirm the robustness of the plasma-polymerised structure under high-rate operation. Impedance spectra recorded before and after cycling are shown in **Figure 6h**. The Nyquist plots are characterized by a semicircle at high frequencies and a straight line at low frequencies. The semicircle is associated with the charge-transfer resistance at the electrode/electrolyte interface, while at low frequencies diffusion process gives rise to a Warburg impedance with a 45° slope. As frequencies decrease, the line evolves into a (purely) capacitive response, shifting the phase angle from 45° to nearly 90°.[60,61] Before and after cycling, the semicircle remains essentially unchanged, while the Warburg line extends to lower frequencies toward the end of the cycling, demonstrating the excellent structural and interfacial stability of the FePc30W@CNFs electrode. Additional impedance data are provided in **Figure S10**.

**Figure 7** shows the XPS analysis of the FePc30W@CNFs electrode after 6000 charge-discharge cycles, revealing clear surface chemical changes. The survey spectrum evidences a strong alteration of the surface composition (C = 85.4 %, N = 5.1 %, O = 6.7 %, Fe = 0.3 %, K = 2.5 %), markedly different from the pristine film (C = 68.9 %, N = 19.0 %, O = 11.1 %, Fe = 1.0 %) (**Figure 7b**). The carbon enrichment and parallel decrease in N and Fe indicate partial loss or detachment of the plasma-polymer coating and exposure of bare fibres, as shown in **Figure S11**. The appearance of K2p features arises from KOH residues retained after rinsing. Despite these changes, the high-resolution C1s+K2p spectrum (**Figure 7f**) reveals that the main carbon functionalities, C=C (284.4 eV), C–C/C–H (284.8 eV), C–N (286.0 eV), N–C=N (286.5 eV), C=O (287.8 eV), O–C=O (289.0 eV), and $\pi$–$\pi$* interactions (291.1 eV) remain well defined, confirming that the carbon framework largely preserves its chemistry. A new component at 290.0 eV indicates the formation of potassium carboxylates (–COOK),[62] and the K2p$_{3/2}$ and K2p$_{1/2}$ peaks at 293.2 and 296.0 eV display the expected 2.8 eV spin-orbit splitting[63]. In the N1s region (**Figure 7e**), the N$_\beta$/N$_\alpha$ ratio increases from 1.15 (pristine) to 2.4 after cycling, revealing a preferential loss of N$_\alpha$ species coordinated to Fe, which are more susceptible to OH⁻ attack in alkaline electrolyte. This selective degradation reduces Fe-N active sites, consistent with the moderate capacitance loss observed after long-term cycling. The Fe2p spectrum (**Figure 7c**) confirms that Fe-N coordination is largely retained, although the $Fe^{3+}/Fe^{2+}$ ratio increases from 0.30 to 0.50 due to mild surface oxidation. This partial but not severe oxidation explains why the capacitance decreases gradually rather than abruptly. Finally, the O1s spectra (**Figure 7d**) show an increase in $O_{Fe}$ and $O_{H2O}$, consistent with Fe oxidation and enhanced electrolyte/water retention. The increase in oxygen vacancy related contributions after cycling likely reflects electrochemically induced surface restructuring and dehydroxylation processes, which generate additional defect sites while preserving most Fe–N coordination. The partial coating detachment discussed above is mainly attributed to mechanical effects during post-cycling handling rather than to the electrochemical process itself. This structural reorganisation accounts for the gradual decline in capacitance observed during long-term cycling.

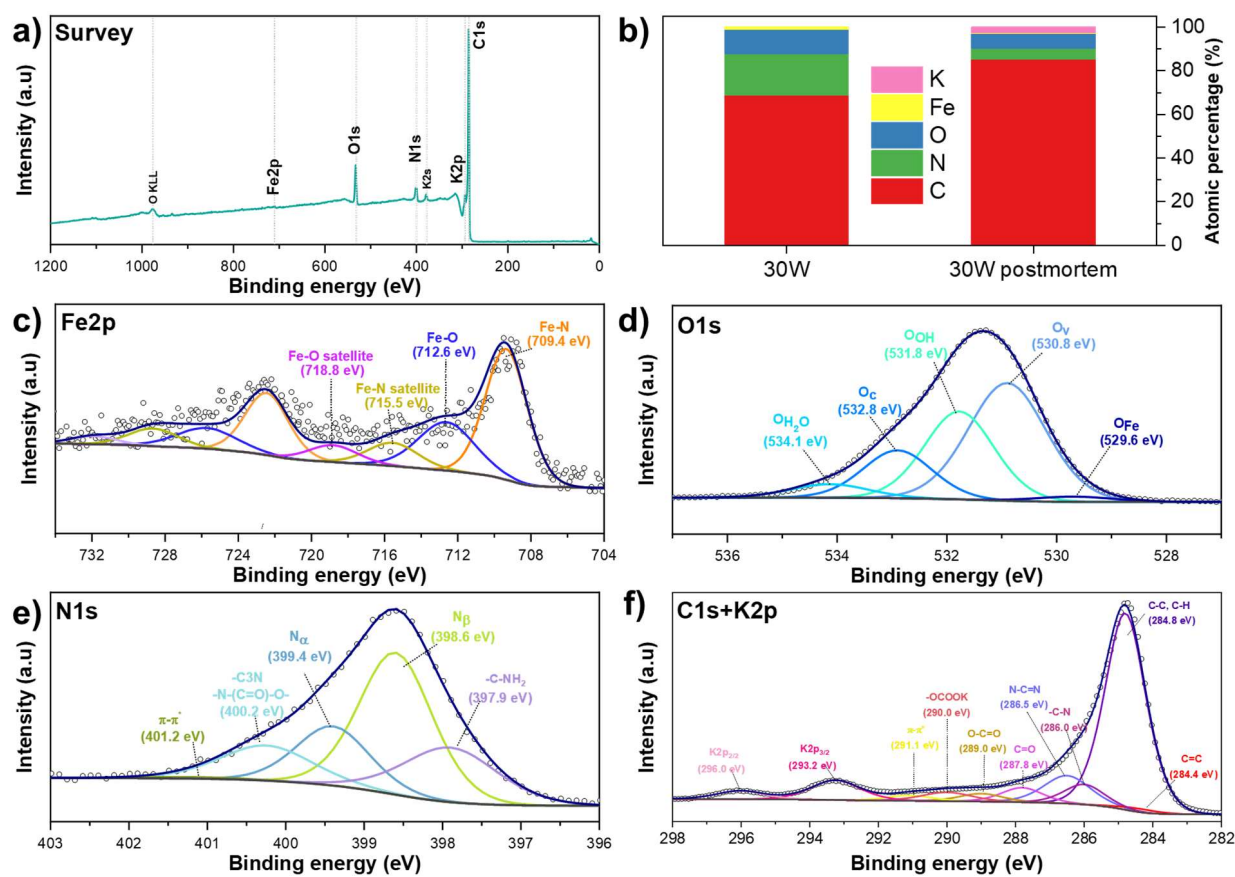

**Figure 7.** (a) XPS survey spectra of the FePc30W@CNFs sample after 6000 charge-discharge cycles. (b) Surface atomic percentages determined by XPS, compared with those of the pristine FePc30W@CNFs film. Detailed high-resolution spectra of (c) Fe2p; (d) O1s; (e) N1s and (f) C1s+K2p regions.

To further investigate surface structural changes during operation, *in-situ* Raman measurements were performed. The experimental conditions and the measurement set-up are described in Section 2 and illustrated in **Figure S12**. **Figure 8** shows the spectra collected during charging (a) and discharging (b). It can be noted that all Raman bands progressively attenuate upon charging, without noticeable shifts in peak position, reaching minimum intensity at 0.75 V. During discharge, the bands fully recover, confirming that the attenuation is a reversible physical effect rather than chemical degradation. The absence of peak shifts or band broadening indicates that no significant structural transformation occurs, and the intensity modulation is instead attributed to reversible changes in the electronic structure under polarization. Variations in charge density and oxidation state can modify the molecular polarizability, thereby affecting the Raman scattering efficiency without altering the vibrational framework. No new Raman features appear, in contrast with previous reports where irreversible oxidation or $Fe^{2+}$ demetalation was observed.[39,64] Notably, the Fe–N stretching band at 594 cm$^{-1}$ remains visible throughout the charge-discharge process, further confirming the stability of the metal-ligand coordination. Overall, the FePc30W@CNFs electrode responds to the applied potential through a reversible reorganization of its chemical environment while preserving structural integrity, fully consistent with the XPS results.

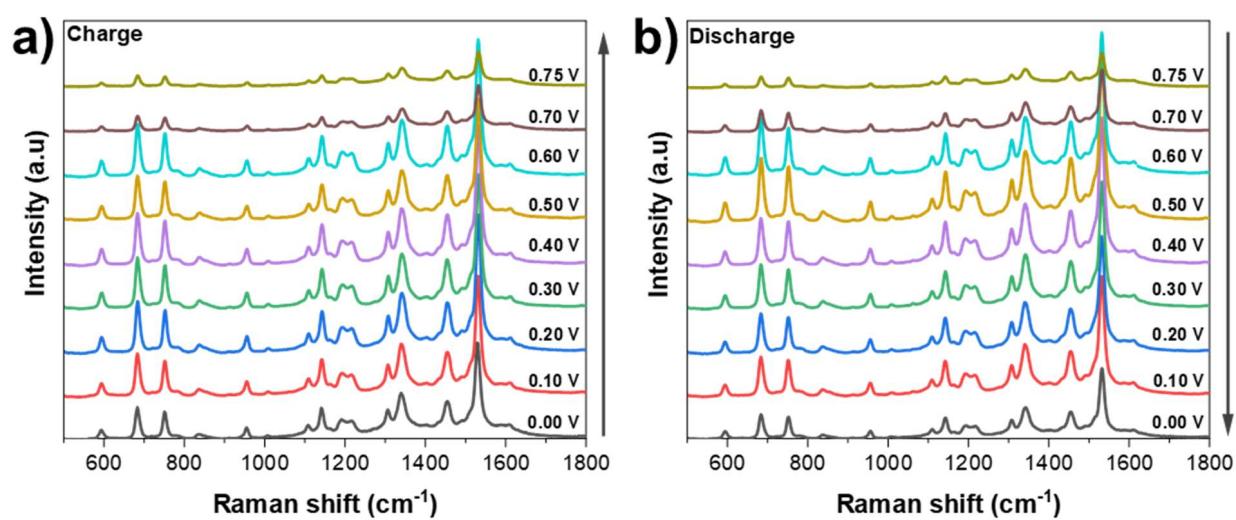

**Figure 8.** *In situ* Raman spectra recorded under electrochemical potentiostatic conditions between 0 V and 0.75 V, during (a) charging and (b) discharging processes.

To benchmark the FePc30W@CNFs electrode against reported MPc-based systems, a Ragone plot was elaborated, which relates energy density (Wh·kg$^{-1}$ or Wh·cm$^{-2}$) to power density (W·kg$^{-1}$ or W·cm$^{-2}$). The energy density of a single electrode is given by:

$$E_{electroded\ (gravimetric)} = \frac{C_{electrode} \cdot \Delta V^2}{2 \cdot 3.6} \qquad (6)$$

$$E_{electroded\ (areal)} = \frac{C_{electrode} \cdot \Delta V^2}{2 \cdot 3600} \qquad (7)$$

where $C_{electrode}$ is the specific (gravimetric or areal) capacitance obtained from GCD curves, and $\Delta V$ the potential window. The constants 3.6 and 3600 convert joules into Wh in gravimetric and areal units, respectively. The corresponding power density is defined as:

$$P_{electrode} = \frac{E_{electrode}}{\Delta t} \cdot 3600 \qquad (8)$$

**Figure 9** compares FePc30W@CNFs with previously reported MPc-based supercapacitors (FePc, CoPc, NiPc, CuPc and ZnPc), using literature-extracted energy/power values (full dataset in **Table S2**).[11,14–16,65–69] The blue star markers corresponding to FePc30W@CNFs were constructed from the energy and power densities calculated from individual galvanostatic charge-discharge (GCD) measurements at different current densities (0.15, 0.25, 0.50 A·g$^{-1}$, etc., see **Figure 6e**); each point therefore represents a discrete experimental value rather than a fitted curve. The FePc30W@CNFs electrode achieves a high energy density of 7.42 Wh·kg$^{-1}$ at 225 W·kg$^{-1}$ (6.2·10$^{-5}$ Wh·cm$^{-2}$ at 3.6·10$^{-4}$ W·cm$^{-2}$), and still delivers 0.85 Wh·kg$^{-1}$ at 6000 W·kg$^{-1}$, demonstrating fast charge-transfer kinetics and excellent rate capability (8.2·10$^{-6}$ Wh·cm$^{-2}$ at 0.06 W·cm$^{-2}$). This Ragone profile highlights a stable energy-power balance across the full operating range

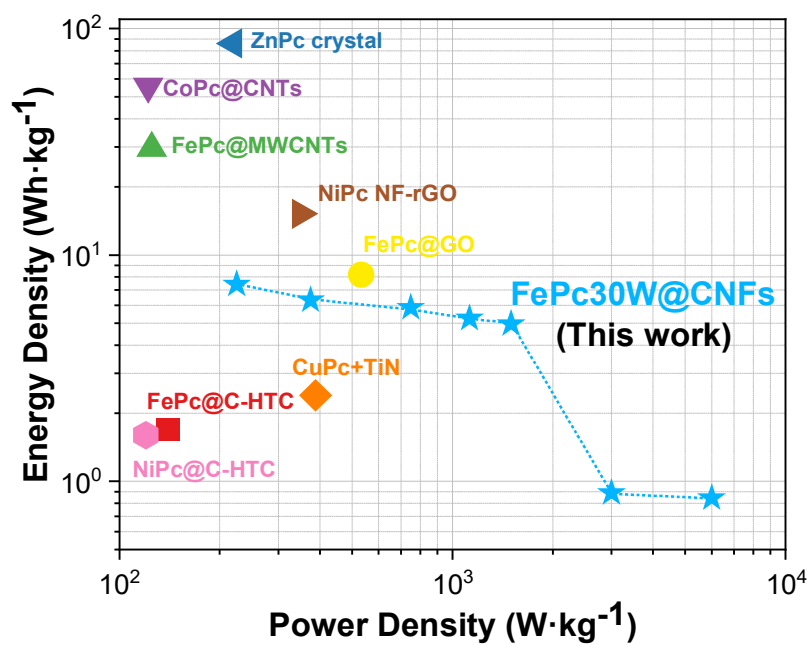

**Figure 9.** Ragone plot comparing FePc30W@CNFs electrode (blue star markers) with previously reported metal-phthalocyanine-based supercapacitors. The FePc30W@CNFs data points were obtained from energy and power densities calculated from individual GCD measurements at different current densities (see **Figure 6e**).[11,14–16,65–69]

These results underscore the advantage of the RPAVD-$N_2$ plasma polymerisation process, which stabilises the Fe–N electroactive centres while producing a nitrogen-rich, conductive and mechanically robust coating on CNFs. This synergistic architecture supports efficient pseudocapacitive reactions alongside rapid ion/electron transport, positioning FePc30W@CNFs as a high-performance, robust supercapacitor and confirming plasma-assisted polymerisation as an effective strategy for energy-storage electrode engineering.

## 4. Conclusions

A novel methodology based on RPAVD of iron (II) phthalocyanine plasma polymer using remote $N_2$ plasma was employed to fabricate FePc@CNTs electrode materials for supercapacitors. The remote $N_2$ plasma plays a dual role: it functionalises the as-prepared CNTs by introducing nitrogen-containing groups and defect sites, and simultaneously enables the plasma-assisted polymerisation of FePc, forming a conformal coating with tuneable chemical composition. The RPAVD process is fully conformal and operates at room temperature, ensuring compatibility with delicate substrates. The resulting plasma polymers exhibit enhanced long-term environmental stability compared to the original FePc molecule, making them highly suitable for durable energy storage applications. Notably, the FePc30W@CNFs system delivers a specific capacitance nearly 9.5 times higher than that of the sublimated FePc@CNFs electrode, a result attributed to the preservation of Fe–N electroactive centres, the incorporation of nitrogen-rich functionalities, and the formation of a conductive and mechanically stable coating that retains structural integrity without demetalation or irreversible degradation under repeated charge-discharge cycling. The assembled symmetric device based on FePc30W@CNFs delivered a high energy density of 7.42 Wh·kg$^{-1}$ at a specific power of 225 W·kg$^{-1}$ and retained 0.85 Wh·kg$^{-1}$ at 6000 W·kg$^{-1}$, with excellent cycling stability (87.5% retention after 6000 cycles). This work demonstrates that remote plasma polymerisation constitutes an effective route to produce electrodes exhibiting high energy and fast power delivery. This work demonstrates RPAVD as a successful strategy for engineering durable phthalocyanines-based supercapacitors.

**References**

(1) Zhang, J.; Gu, M.; Chen, X. Supercapacitors for Renewable Energy Applications: A Review. *Micro and Nano Engineering* **2023**, *21*, 100229. https://doi.org/10.1016/j.mne.2023.100229.

(2) Dissanayake, K.; Kularatna-Abeywardana, D. A Review of Supercapacitors: Materials, Technology, Challenges, and Renewable Energy Applications. *Journal of Energy Storage* **2024**, *96*, 112563. https://doi.org/10.1016/j.est.2024.112563.

(3) Singh, N.; Singh, V.; Bisht, N.; Negi, P.; Dhyani, A.; Sharma, R. K.; Tewari, B. S. A Comprehensive Review on Supercapacitors: Basics to Recent Advancements. *Journal of Energy Storage* **2025**, *121*, 116498. https://doi.org/10.1016/j.est.2025.116498.

(4) Vessally, E.; Rzayev, R. M.; Niyazova, A. A.; Aggarwal, T.; Rahimova, K. E. Overview of Recent Developments in Carbon-Based Nanocomposites for Supercapacitor Applications. *RSC Adv.* **2024**, *14* (54), 40141–40159. https://doi.org/10.1039/D4RA08446B.

(5) Ahmad, F.; Shahzad, A.; Danish, M.; Fatima, M.; Adnan, M.; Atiq, S.; Asim, M.; Khan, M. A.; Ain, Q. U.; Perveen, R. Recent Developments in Transition Metal Oxide-Based Electrode Composites for Supercapacitor Applications. *Journal of Energy Storage* **2024**, *81*, 110430. https://doi.org/10.1016/j.est.2024.110430.

(6) Vinayak, V. J. V.; Deshmukh, K.; Murthy, V. R. K.; Pasha, S. K. K. Conducting Polymer Based Nanocomposites for Supercapacitor Applications: A Review of Recent Advances, Challenges and Future Prospects. *Journal of Energy Storage* **2024**, *100*, 113551. https://doi.org/10.1016/j.est.2024.113551.

(7) Ding, W.; Xiao, L.; Lv, L.-P.; Wang, Y. Redox-Active Organic Electrode Materials for Supercapacitors. *Batteries & Supercaps* **2023**, *6* (11), e202300278. https://doi.org/10.1002/batt.202300278.

(8) Alharbi, A. Boosting Supercapacitor Performance through Innovative Transition Metal-Based Electrode Materials. *RSC Adv.* **2025**, *15* (41), 34551–34582. https://doi.org/10.1039/D5RA02905H.

(9) Mahala, S.; Khosravinia, K.; Kiani, A. Unwanted Degradation in Pseudocapacitors: Challenges and Opportunities. *Journal of Energy Storage* **2023**, *67*, 107558. https://doi.org/10.1016/j.est.2023.107558.

(10) Rahat, S. M. S. M.; Hasan, K. M. Z.; Mondol, M. M. H.; Mallik, A. K. A Comprehensive Review of Carbon Nanotube-Based Metal Oxide Nanocomposites for Supercapacitors. *Journal of Energy Storage* **2023**, *73*, 108847. https://doi.org/10.1016/j.est.2023.108847.

(11) Balogun, S. A.; Mokethe, S.; Masekela, D.; Thole, D.; Teffu, D. M.; Moronkola, A. A.; Makgopa, K.; Modibane, K. D. Engineering CoPcMWCNTs and NiPcMWCNTs Nanocomposites for Photocatalytic Hydrogen Production and Supercapacitor Applications. *International Journal of Hydrogen Energy* **2025**, *176*, 151494. https://doi.org/10.1016/j.ijhydene.2025.151494.

(12) Oni, J.; Ozoemena, K. I. Phthalocyanines in Batteries and Supercapacitors. *Journal of Porphyrins and Phthalocyanines* **2012**, *16* (07n08), 754–760. https://doi.org/10.1142/S1088424612300078.

(13) Gorduk, O.; Gencten, M.; Gorduk, S.; Sahin, M.; Sahin, Y. Electrochemical Fabrication and Supercapacitor Performances of Metallo Phthalocyanine/Functionalized-Multiwalled Carbon Nanotube/Polyaniline Modified Hybrid Electrode Materials. *Journal of Energy Storage* **2021**, *33*, 102049. https://doi.org/10.1016/j.est.2020.102049.

(14) Deyab, M. A.; Mohsen, Q.; Slavcheva, E. Co-Phthalocyanin/CNTs Nanocomposites: Synthesis, Characterizations, and Application as an Efficient Supercapacitor. *Journal of Molecular Liquids* **2022**, *359*, 119319. https://doi.org/10.1016/j.molliq.2022.119319.

(15)   Sanchez-Sanchez, A.; Izquierdo, M. T.; Mathieu, S.; Ghanbaja, J.; Celzard, A.; Fierro, V. Structure and Electrochemical Properties of Carbon Nanostructures Derived from Nickel(II) and Iron(II) Phthalocyanines. *Journal of Advanced Research* **2020**, *22*, 85–97. https://doi.org/10.1016/j.jare.2019.11.004.

(16)   Wang, Q.; Gao, H.; Zhao, C.; Wang, S.; Liu, X.; Wang, Z.; Yu, J.; Kwon, Y.-U.; Zhao, Y. π-π Stacked Iron (II) Phthalocyanine/Graphene Oxide Composites: Rational Fabrication and Excellent Supercapacitor Properties with Superior Rate Performance. *Journal of Solid State Electrochemistry* **2021**, *25* (2), 659–670. https://doi.org/10.1007/s10008-020-04840-7.

(17)   Khan, T. M. Chapter 15 - Plasma Technology for Supercapacitors. In *Energy From Plasma*; Yasin, G., Nguyen, D. B., Gupta, R. K., Ajmal, S., Nguyen, T. A., Eds.; Woodhead Publishing, 2025; pp 433–463. https://doi.org/10.1016/B978-0-443-26584-6.00015-4.

(18)   Liang, X.; Liu, P.; Qiu, Z.; Shen, S.; Cao, F.; Zhang, Y.; Chen, M.; He, X.; Xia, Y.; Wang, C.; Wan, W.; Zhang, J.; Huang, H.; Gan, Y.; Xia, X.; Zhang, W. Plasma Technology for Advanced Electrochemical Energy Storage. *Chemistry – A European Journal* **2024**, *30* (19), e202304168. https://doi.org/10.1002/chem.202304168.

(19)   Sahoo, S.; Sahoo, G.; Jeong, S. M.; Rout, C. S. A Review on Supercapacitors Based on Plasma Enhanced Chemical Vapor Deposited Vertical Graphene Arrays. *Journal of Energy Storage* **2022**, *53*, 105212. https://doi.org/10.1016/j.est.2022.105212.

(20)   Liu, F.; Zhang, L.-H.; Zhang, Z.; Zhou, Y.; Zhang, Y.; Huang, J.-L.; Fang, Z. The Application of Plasma Technology for the Preparation of Supercapacitor Electrode Materials. *Dalton Trans.* **2024**, *53* (13), 5749–5769. https://doi.org/10.1039/D3DT04362B.

(21)   Liu, F.; Zhang, L.-H.; Zhang, Z.; Zhou, Y.; Zhang, Y.; Huang, J.-L.; Fang, Z. The Application of Plasma Technology for the Preparation of Supercapacitor Electrode Materials. *Dalton Trans.* **2024**, *53* (13), 5749–5769. https://doi.org/10.1039/D3DT04362B.

(22)   Obrero-Perez, J. M.; Contreras-Bernal, L.; Nuñez-Galvez, F.; Castillo-Seoane, J.; Valadez-Villalobos, K.; Aparicio, F. J.; Anta, J. A.; Borras, A.; Sanchez-Valencia, J. R.; Barranco, A. Ultrathin Plasma Polymer Passivation of Perovskite Solar Cells for Improved Stability and Reproducibility. *Advanced Energy Materials* **2022**, *12* (32), 2200812. https://doi.org/10.1002/aenm.202200812.

(23)   Obrero, J. M.; Moreno-Martinez, G. P.; Rojas, T. C.; Ferrer, F. J.; Moscoso, F. G.; Contreras-Bernal, L.; Castillo-Seoane, J.; Nuñez-Galvez, F.; Aparicio Rebollo, F. J.; Borras, A.; Sanchez-Valencia, J. R.; Barranco, A. Enhanced Luminous Transmission and Solar Modulation in Thermochromic VO2 Aerogel-like Films via Remote Plasma Deposition. *ACS Appl. Mater. Interfaces* **2025**, *17* (39), 55172–55188. https://doi.org/10.1021/acsami.5c07264.

(24)   Obrero, J. M.; Contreras-Bernal, L.; Aparicio Rebollo, F. J.; Rojas, T. C.; Ferrer, F. J.; Orozco, N.; Saghi, Z.; Czermak, T.; Pedrosa, J. M.; López-Santos, C.; Ostrikov, K. K.; Borras, A.; Sánchez-Valencia, J. R.; Barranco, A. Conformal TiO2 Aerogel-Like Films by Plasma Deposition: From Omniphobic Antireflective Coatings to Perovskite Solar Cell Photoelectrodes. *ACS Appl. Mater. Interfaces* **2024**, *16* (30), 39745–39760. https://doi.org/10.1021/acsami.4c00555.

(25)   Obrero-Perez, J. M.; Nuñez-Galvez, F.; Contreras-Bernal, L.; Castillo-Seoane, J.; Moreno, G. P.; Czermak, T.; Aparicio, F. J.; Rojas, T. C.; Ferrer, F. J.; Borras, A.; Barranco, A.; Sánchez-Valencia, J. R. Low-Temperature Remote Plasma Synthesis of Highly Porous TiO2 as Electron Transport Layers in Perovskite Solar Cells. *Advanced Materials Interfaces* **2025**, *12* (21), e00241. https://doi.org/10.1002/admi.202500241.

(26)    Nabil, M.; Contreras-Bernal, L.; Moreno-Martinez, G. P.; Obrero-Perez, J.; Castillo-Seoane, J.; Anta, J. A.; Oskam, G.; Pistor, P.; Borrás, A.; Sanchez-Valencia, J. R.; Barranco, A. Boosting Perovskite Solar Cell Stability: Dual Protection with Ultrathin Plasma Polymer Passivation Layers. *Materials Today Energy* **2025**, *54*, 102117. https://doi.org/10.1016/j.mtener.2025.102117.

(27)    Aparicio, F. J.; Holgado, M.; Borras, A.; Blaszczyk-Lezak, I.; Griol, A.; Barrios, C. A.; Casquel, R.; Sanza, F. J.; Sohlström, H.; Antelius, M.; González-Elipe, A. R.; Barranco, A. Transparent Nanometric Organic Luminescent Films as UV-Active Components in Photonic Structures. *Adv. Mater.* **2011**, *23* (6), 761–765. https://doi.org/10.1002/adma.201003088.

(28)    Biesinger, M. C.; Payne, B. P.; Grosvenor, A. P.; Lau, L. W. M.; Gerson, A. R.; Smart, R. S. C. Resolving Surface Chemical States in XPS Analysis of First Row Transition Metals, Oxides and Hydroxides: Cr, Mn, Fe, Co and Ni. *Appl. Surf. Sci.* **2011**, *257* (7), 2717–2730. https://doi.org/10.1016/j.apsusc.2010.10.051.

(29)    Grey, L. H.; Nie, H.-Y.; Biesinger, M. C. Defining the Nature of Adventitious Carbon and Improving Its Merit as a Charge Correction Reference for XPS. *Applied Surface Science* **2024**, *653*, 159319. https://doi.org/10.1016/j.apsusc.2024.159319.

(30)    Alcaire, M.; Aparicio, F. J.; Obrero, J.; López-Santos, C.; Garcia-Garcia, F. J.; Sánchez-Valencia, J. R.; Frutos, F.; Ostrikov, K.; Borrás, A.; Barranco, A. Plasma Enabled Conformal and Damage Free Encapsulation of Fragile Molecular Matter: From Surface-Supported to On-Device Nanostructures. *Adv. Funct. Mater.*, 2019, *29*. https://doi.org/10.1002/adfm.201903535.

(31)    Obrero, J. M.; Filippin, A. N.; Alcaire, M.; Sanchez-Valencia, J. R.; Jacob, M.; Matei, C.; Aparicio, F. J.; Macias-Montero, M.; Rojas, T. C.; Espinos, J. P.; Saghi, Z.; Barranco, A.; Borras, A. Supported Porous Nanostructures Developed by Plasma Processing of Metal Phthalocyanines and Porphyrins. *Frontiers in Chemistry* **2020**, *8*. https://doi.org/10.3389/fchem.2020.00520.

(32)    Zanguina, A.; Bayo-Bangoura, M.; Bayo, K.; Ouedraogo, G. V. IR AND UV-VISIBLE SPECTRA OF IRON(II) PHTHALOCYANINE COMPLEXES WITH PHOSPHINE OR PHOSPHITE. *Bull. Chem. Soc. Eth.* **2002**, *16* (1), 73–79. https://doi.org/10.4314/bcse.v16i1.20950.

(33)    Islam, T.; Helsel, N.; Choudhury, P. Facile Synthesis of Iron Phthalocyanine Functionalized B-Doped Graphene Composite Using a Green Solvent for Superior ORR Performance. *Molecular Catalysis* **2025**, *587*, 115500. https://doi.org/10.1016/j.mcat.2025.115500.

(34)    Barranco, A.; Aparicio, F.; Yanguas-Gil, A.; Groening, P.; Cotrino, J.; Ganzález-Elipe, A. R. Optically Active Thin Films Deposited by Plasma Polymerization of Dye Molecules. *Chemical Vapor Deposition*, 2007, *13*, 319–325. https://doi.org/10.1002/cvde.200606552.

(35)    Myers, J. F.; Canham, G. W. Rayner.; Lever, A. B. P. Higher Oxidation Level Phthalocyanine Complexes of Chromium, Iron, Cobalt and Zinc.  Phthalocyanine Radical Species. *Inorg. Chem.* **1975**, *14* (3), 461–468. https://doi.org/10.1021/ic50145a002.

(36)    Ziegler, C. J.; Nemykin, V. N. The Fascinating Story of Axial Ligand Dependent Spectroscopy and Redox-Properties in Iron(Ii) Phthalocyanines. *Dalton Trans.* **2023**, *52* (43), 15647–15655. https://doi.org/10.1039/D3DT02565A.

(37)    Thielke, M. W.; Lopez Guzman, S.; Victoria Tafoya, J. P.; García Tamayo, E.; Castro Herazo, C. I.; Hosseinaei, O.; Sobrido, A. J. Full Lignin-Derived Electrospun Carbon Materials as Electrodes for Supercapacitors. *Frontiers in Materials* **2022**, *Volume 9-2022*. https://doi.org/10.3389/fmats.2022.859872.

(38)    Liu, X.; Choi, J.; Xu, Z.; Grey, C. P.; Fleischmann, S.; Forse, A. C. Raman Spectroscopy Measurements Support Disorder-Driven Capacitance in Nanoporous

Carbons. *J. Am. Chem. Soc.* **2024**, *146* (45), 30748–30752. https://doi.org/10.1021/jacs.4c10214.

(39) Chen, Z.; Jiang, S.; Kang, G.; Nguyen, D.; Schatz, G. C.; Van Duyne, R. P. Operando Characterization of Iron Phthalocyanine Deactivation during Oxygen Reduction Reaction Using Electrochemical Tip-Enhanced Raman Spectroscopy. *J. Am. Chem. Soc.* **2019**, *141* (39), 15684–15692. https://doi.org/10.1021/jacs.9b07979.

(40) Liu, Z.; Zhang, X.; Zhang, Y.; Jiang, J. Theoretical Investigation of the Molecular, Electronic Structures and Vibrational Spectra of a Series of First Transition Metal Phthalocyanines. *Spectrochimica Acta Part A: Molecular and Biomolecular Spectroscopy* **2007**, *67* (5), 1232–1246. https://doi.org/10.1016/j.saa.2006.10.013.

(41) Tackley, D. R.; Dent, G.; Ewen Smith, W. Phthalocyanines: Structure and Vibrations. *Phys. Chem. Chem. Phys.* **2001**, *3* (8), 1419–1426. https://doi.org/10.1039/B007763L.

(42) Szybowicz, M.; Makowiecki, J. Orientation Study of Iron Phthalocyanine (FePc) Thin Films Deposited on Silicon Substrate Investigated by Atomic Force Microscopy and Micro-Raman Spectroscopy. *Journal of Materials Science* **2012**, *47* (3), 1522–1530. https://doi.org/10.1007/s10853-011-5940-7.

(43) Hellgren, N.; Haasch, R. T.; Schmidt, S.; Hultman, L.; Petrov, I. Interpretation of X-Ray Photoelectron Spectra of Carbon-Nitride Thin Films: New Insights from in Situ XPS. *Carbon* **2016**, *108*, 242–252. https://doi.org/10.1016/j.carbon.2016.07.017.

(44) Komba, N.; Zhang, G.; Wei, Q.; Yang, X.; Prakash, J.; Chenitz, R.; Rosei, F.; Sun, S. Iron (II) Phthalocyanine/N-Doped Graphene: A Highly Efficient Non-Precious Metal Catalyst for Oxygen Reduction. *International Journal of Hydrogen Energy* **2019**, *44* (33), 18103–18114. https://doi.org/10.1016/j.ijhydene.2019.05.032.

(45) Chen, K.; Liu, K.; An, P.; Li, H.; Lin, Y.; Hu, J.; Jia, C.; Fu, J.; Li, H.; Liu, H.; Lin, Z.; Li, W.; Li, J.; Lu, Y.-R.; Chan, T.-S.; Zhang, N.; Liu, M. Iron Phthalocyanine with Coordination Induced Electronic Localization to Boost Oxygen Reduction Reaction. *Nature Communications* **2020**, *11* (1), 4173. https://doi.org/10.1038/s41467-020-18062-y.

(46) Sun, Q.; Wang, Z.; Zhou, M.; Li, J.; Lu, R.; Wang, Y.; Liao, X.; Zhao, Y. Tailoring Activity of Iron Phthalocyanine by Edge-Nitrogen Sites Induced Electronic Delocalization. *Applied Surface Science* **2023**, *624*, 157154. https://doi.org/10.1016/j.apsusc.2023.157154.

(47) Sahu, R. K.; Mukherjee, D.; Tiwari, J. P.; Mishra, T.; Roy, S. K.; Pathak, L. C. Influence of Foreign Fe Ions on Wet Chemical Synthesis of Pt Nanoparticle Thin Films at Ambient Temperature: In Situversus Direct Addition. *J. Mater. Chem.* **2009**, *19* (37), 6810–6815. https://doi.org/10.1039/B908080E.

(48) Paul, A.; Ghosh, S.; Kolya, H.; Kang, C.-W.; Murmu, N. C.; Kuila, T. New Insight into the Effect of Oxygen Vacancies on Electrochemical Performance of Nickel-Tin Oxide/Reduced Graphene Oxide Composite for Asymmetric Supercapacitor. *Journal of Energy Storage* **2023**, *62*, 106922. https://doi.org/10.1016/j.est.2023.106922.

(49) Wang, H.; Mi, N.; Sun, S.; Zhang, W.; Yao, S. Oxygen Vacancies Enhancing Capacitance of MgCo2O4 for High Performance Asymmetric Supercapacitors. *Journal of Alloys and Compounds* **2021**, *869*, 159294. https://doi.org/10.1016/j.jallcom.2021.159294.

(50) Marshall-Roth, T.; Libretto, N. J.; Wrobel, A. T.; Anderton, K. J.; Pegis, M. L.; Ricke, N. D.; Voorhis, T. V.; Miller, J. T.; Surendranath, Y. A Pyridinic Fe-N4 Macrocycle Models the Active Sites in Fe/N-Doped Carbon Electrocatalysts. *Nat Commun* **2020**, *11* (1), 5283. https://doi.org/10.1038/s41467-020-18969-6.

(51) Olayo, M. G.; Alvarado, E. J.; González-Torres, M.; Gómez, L. M.; Cruz, G. J. Quantifying Amines in Polymers by XPS. *Polymer Bulletin* **2024**, *81* (3), 2319–2328. https://doi.org/10.1007/s00289-023-04829-y.

(52)     Kanani-Jazi, M. H.; Akbari, S. Quantitative XPS Analysis of Amine-Terminated Dendritic Functionalized Halloysite Nanotubes Decorated on PAN Nanofibrous Membrane and Adsorption/Filtration of Cr(VI). *Chemical Engineering Journal* **2024**, *482*, 148746. https://doi.org/10.1016/j.cej.2024.148746.

(53)     Kehrer, M.; Duchoslav, J.; Hinterreiter, A.; Cobet, M.; Mehic, A.; Stehrer, T.; Stifter, D. XPS Investigation on the Reactivity of Surface Imine Groups with TFAA. *Plasma Processes and Polymers* **2019**, *16* (4), 1800160. https://doi.org/10.1002/ppap.201800160.

(54)     Giesbers, M.; Marcelis, A. T. M.; Zuilhof, H. Simulation of XPS C1s Spectra of Organic Monolayers by Quantum Chemical Methods. *Langmuir* **2013**, *29* (15), 4782–4788. https://doi.org/10.1021/la400445s.

(55)     Zhang, P.; Wang, M.; Liu, Y.; Fu, Y.; Gao, M.; Wang, G.; Wang, F.; Wang, Z.; Chen, G.; Yang, S.; Liu, Y.; Dong, R.; Yu, M.; Lu, X.; Feng, X. Largely Pseudocapacitive Two-Dimensional Conjugated Metal–Organic Framework Anodes with Lowest Unoccupied Molecular Orbital Localized in Nickel-Bis(Dithiolene) Linkages. *J. Am. Chem. Soc.* **2023**, *145* (11), 6247–6256. https://doi.org/10.1021/jacs.2c12684.

(56)     Raavi, R.; Archana, S.; Adinarayana Reddy, P.; Elumalai, P. Performances of Dual Carbon Multi-Ion Supercapacitors in Aqueous and Non-Aqueous Electrolytes. *Energy Adv.* **2023**, *2* (3), 385–397. https://doi.org/10.1039/D2YA00271J.

(57)     Sun, Y.; Xu, D.; He, Z.; Zhang, Z.; Fan, L.; Wang, S. Green Fabrication of Pore-Modulated Carbon Aerogels Using a Biological Template for High-Energy Density Supercapacitors. *J. Mater. Chem. A* **2023**, *11* (37), 20011–20020. https://doi.org/10.1039/D3TA04486F.

(58)     Wei, X.; Qiu, B.; Tian, H.; Lv, Y.; Zhang, W.; Qin, Q.; Liu, Z.; Wei, F. Co-Precipitation Reaction: A Facile Strategy for Designing Hierarchical Porous Carbon Nanosheets for EDLCs and Zinc-Ion Hybrid Supercapacitors. *Applied Surface Science* **2023**, *615*, 156280. https://doi.org/10.1016/j.apsusc.2022.156280.

(59)     Wu, Q.; He, T.; Zhang, Y.; Zhang, J.; Wang, Z.; Liu, Y.; Zhao, L.; Wu, Y.; Ran, F. Cyclic Stability of Supercapacitors: Materials, Energy Storage Mechanism, Test Methods, and Device. *J. Mater. Chem. A* **2021**, *9* (43), 24094–24147. https://doi.org/10.1039/D1TA06815F.

(60)     Lazanas, A. Ch.; Prodromidis, M. I. Electrochemical Impedance Spectroscopy—A Tutorial. *ACS Meas. Sci. Au* **2023**, *3* (3), 162–193. https://doi.org/10.1021/acsmeasuresciau.2c00070.

(61)     Iqbal, M. Z.; Faisal, M. M.; Ali, S. R. Integration of Supercapacitors and Batteries towards High-Performance Hybrid Energy Storage Devices. *International Journal of Energy Research* **2021**, *45* (2), 1449–1479. https://doi.org/10.1002/er.5954.

(62)     Caracciolo, L.; Madec, L.; Martinez, H. XPS Analysis of K-Based Reference Compounds to Allow Reliable Studies of Solid Electrolyte Interphase in K-Ion Batteries. *ACS Appl. Energy Mater.* **2021**, *4* (10), 11693–11699. https://doi.org/10.1021/acsaem.1c02400.

(63)     Chang, L.-L.; Hu, C.; Huang, C.-C.; Lin, C.-Y.; Chung, P.-W.; Tung, K.-L. Potassium Citrate-Activated Porous Carbon Nanostructures for CO2 Adsorption and Electroreduction. *ACS Appl. Nano Mater.* **2023**, *6* (10), 8839–8848. https://doi.org/10.1021/acsanm.3c01231.

(64)     Melendres, C. A.; Rios, C. B.; Feng, X.; McMasters, R. In Situ Laser Raman Spectra of Iron Phthalocyanine Adsorbed on Copper and Gold Electrodes. *J. Phys. Chem.* **1983**, *87* (18), 3526–3531. https://doi.org/10.1021/j100241a033.

(65)     Lu, Y.; Pang, X.; Li, M.; Liang, M.; Wang, W.; He, Q.; Qahramon Zarifzoda, A.; Chen, F. In-Situ Preparation of Iron(II)-Phthalocyanine@multi-Walled-CNTs Nanocomposite for Quasi-Solid-State Flexible Symmetric Supercapacitors with Long

Cycling Life. *ChemSusChem* **2025**, *18* (5), e202401940.
https://doi.org/10.1002/cssc.202401940.

(66)    Dong, A.; Lin, Y.; Guo, Y.; Chen, D.; Wang, X.; Ge, Y.; Li, Q.; Qian, J.
Immobilization of Iron Phthalocyanine on MOF-Derived N-Doped Carbon for Promoting
Oxygen Reduction in Zinc-Air Battery. *Journal of Colloid and Interface Science* **2023**,
*650*, 2056–2064. https://doi.org/10.1016/j.jcis.2023.06.043.

(67)    Chattanahalli Devendrachari, M.; Shimoga, G.; Lee, S.-H.; Heo, Y.-H.; Makri
Nimbegondi Kotresh, H.; Palem, R. R.; Kim, S.-Y.; Choi, D.-S. Advancing Energy
Storage Competence through Copper Phthalocyanine-Stabilized Titanium Nitride Hybrid
Nanocomposites for Symmetric Supercapacitors. *ACS Appl. Energy Mater.* **2023**, *6* (21),
11199–11211. https://doi.org/10.1021/acsaem.3c02093.

(68)    Han, D.; Wang, W.; Yu, S.; Qi, W.; Ling, R.; Yang, C.; Liu, G. Stable β-Form Zinc
Phthalocyanine Cathodes for Flexible Zn-Ion Hybrid Supercapacitors with Ultra-Long
Cycling Life. *Chemical Engineering Journal* **2023**, *468*, 143875.
https://doi.org/10.1016/j.cej.2023.143875.

(69)    Madhuri, K. P.; John, N. S. Supercapacitor Application of Nickel Phthalocyanine
Nanofibres and Its Composite with Reduced Graphene Oxide. *Applied Surface Science*
**2018**, *449*, 528–536. https://doi.org/10.1016/j.apsusc.2017.12.021.

Acknowledgements

We thank the projects PID2022-143120OB-I00 and PCI2024-153451 funded by
MCIN/AEI/10.13039/501100011033 and by "ERDF (FEDER) a way of making Europe, the
Fondos NextgenerationEU and Plan de Recuperación, Transformación y Resiliencia". Project
ANGSTROM was selected in the Joint Transnational Call 2023 of ERA.NET" title="http://M-
ERA.NET>M-ERA.NET 3, which is an EU-funded network of about 49 funding organisations
(Horizon 2020 grant agreement No 958174). M. T. acknowledges the funding provided by UK
Research and Innovation for EPSRC post-doctoral fellowship no. UKRI913. A. J. S.
acknowledges the UK Research and Innovation for Future Leaders Fellowship no.
MR/T041412/1.

# Supporting Information

**Remote Plasma Polymers of Iron (II) Phthalocyanine in Polyacrylonitrile-Derived Carbon Electrospun Fibers as Electrode for Supercapacitors.**

Jose M. Obrero [a*], Jorge PV Tafoya [b], Michael Thielke [b], G.P. Moreno-Martínez [a], Lidia Contreras-Bernal [a,c], Jose Ferreira de Sousa Jr [a], Juan Ramón Sánchez-Valencia [a], Angel Barranco [a], and Ana B. Jorge Sobrido [b*].

a) Nanotechnology on Surfaces and Plasma Laboratory, Materials Science Institute of Seville (CSIC-US), C/ Américo Vespucio 49, 41092, Seville, Spain.

b) Centre for Sustainable Engineering, School of Engineering and Materials Science, Faculty of Science and Engineering, Queen Mary University of London, Mile End Road, London, E1 4NS, UK

c) Química-Física. Department of Physical Chemistry, University of Seville, C/Professor García González n° 2, Seville 41012, Spain

**Before plasma N$_2$ pretreatment**     **After plasma N$_2$ pretreatment**

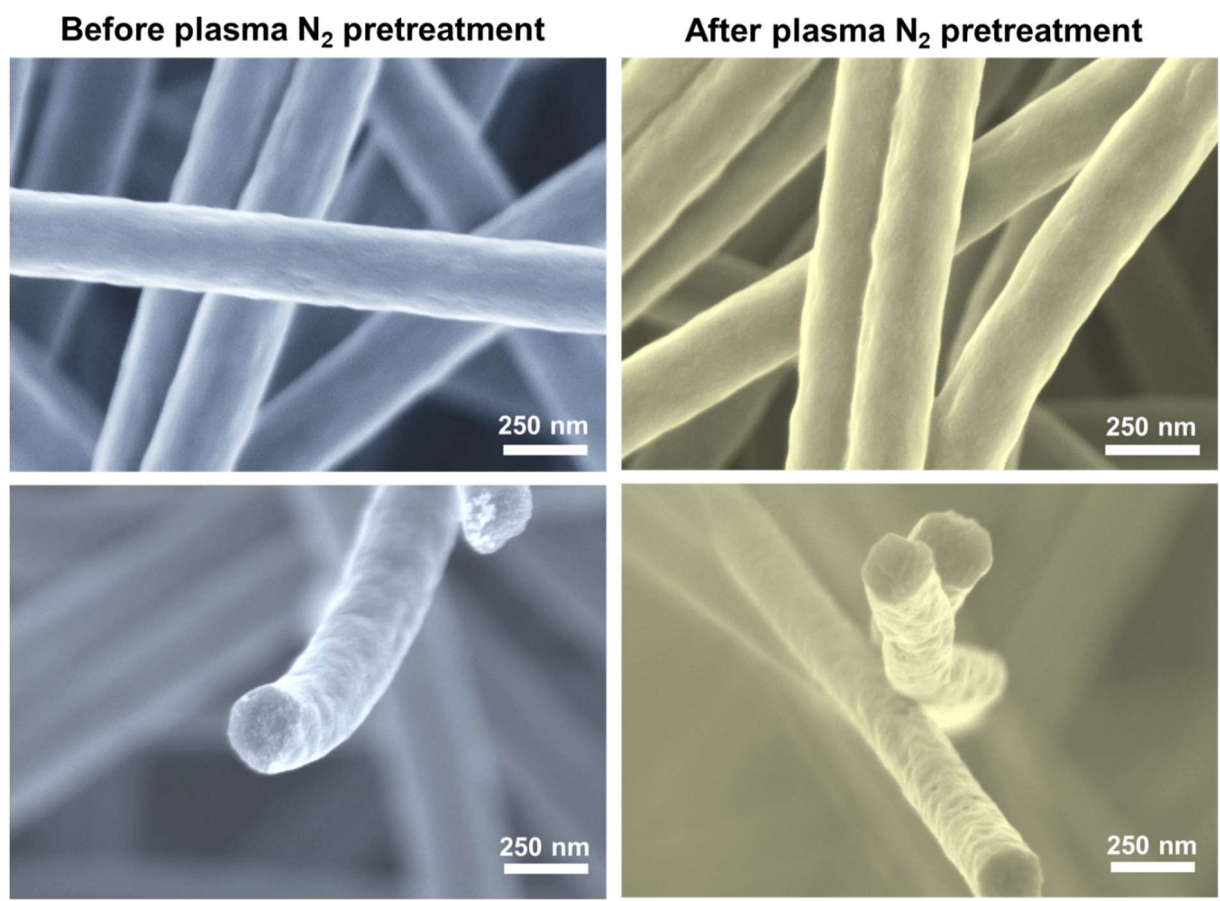

**Figure S1.** Comparative SEM micrographs of carbon nanofibers (CNFs) before and after N$_2$ plasma treatment. The untreated CNFs exhibit smooth and clean surfaces, characteristic of electrospun and carbonized PAN fibers. After exposure to N$_2$ plasma, the overall morphology and fibre diameter remain unchanged, confirming that the process is mild and does not cause structural damage. However, a slight increase in surface roughness can be observed, indicating the introduction of surface defects and nitrogen-containing functional groups that enhance the surface reactivity and adhesion for subsequent FePc deposition.

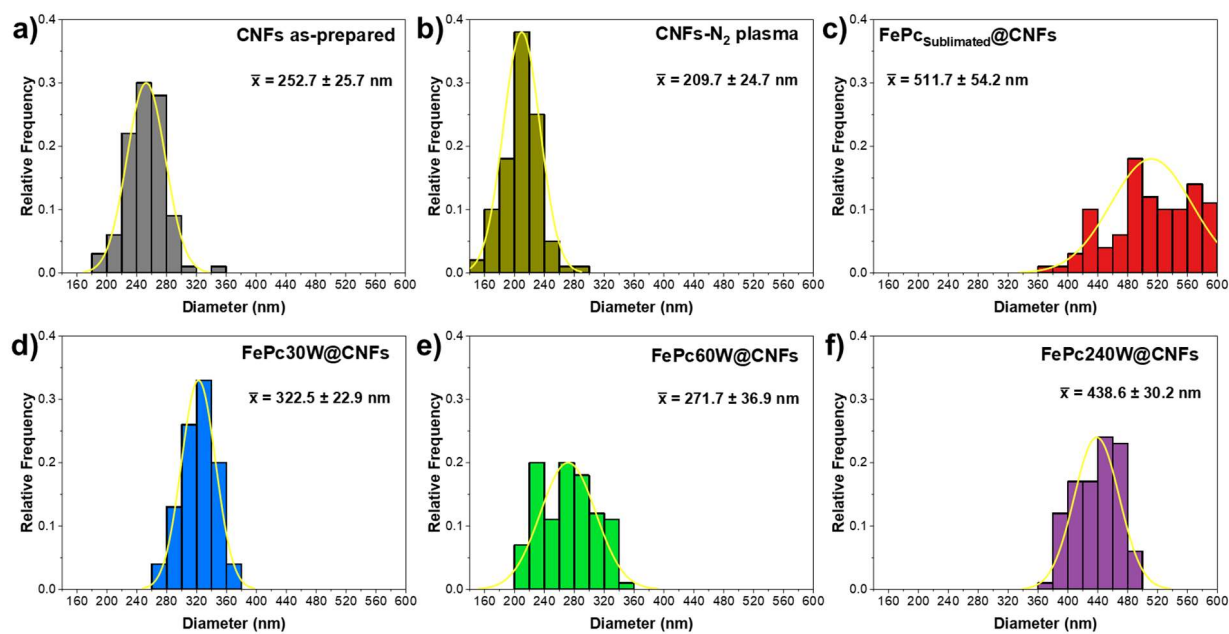

**Figure S2.** Diameter size distribution of CNFs samples determined from the analysis of SEM micrographs of a) CNFs as-prepared, b) CNFs after $N_2$ plasma treatment, c) FePc$_{Sublimated}$@CNFs, d) FePc30W@CNFs, e) FePc60W@CNFs and f) FePc240W@CNFs.

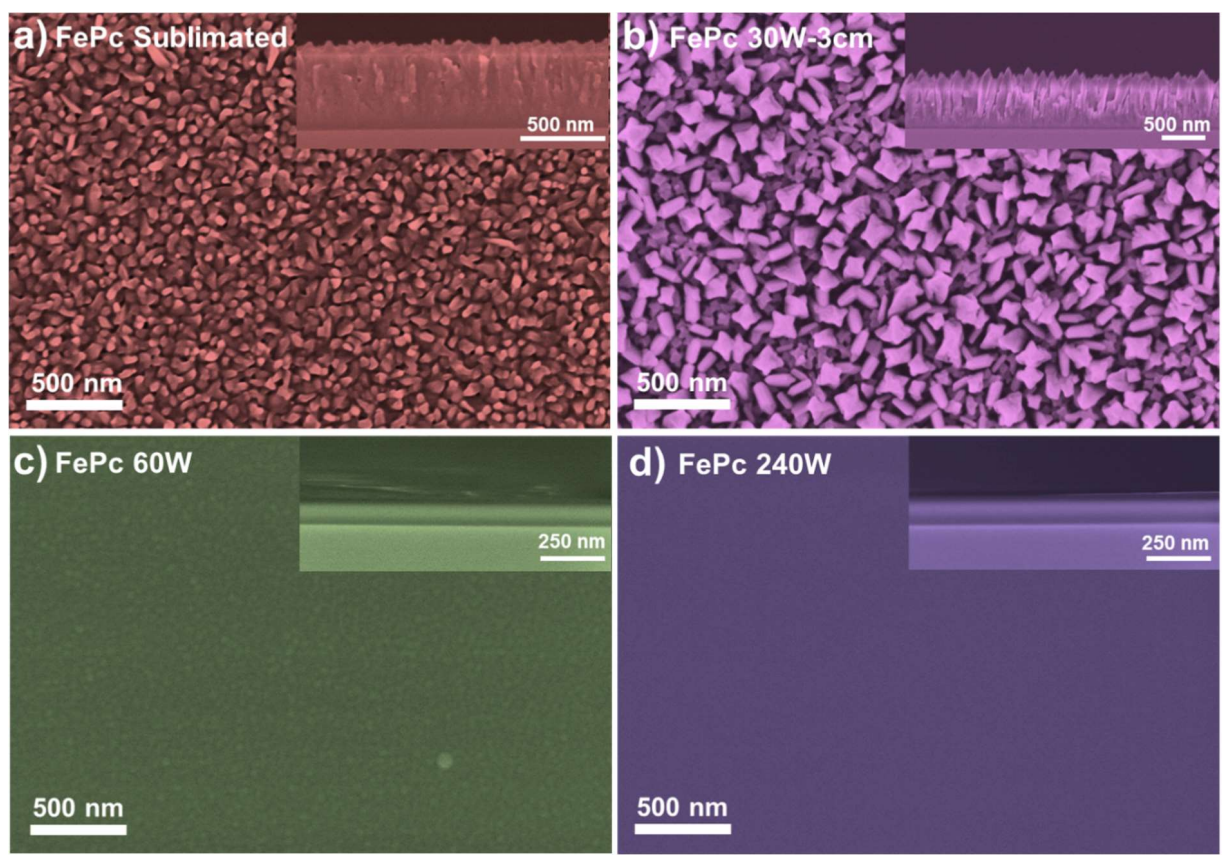

**Figure S3.** Top-view SEM micrographs and cross-sectional (inset) SEM images of FePc films on silicon: sublimated FePc and FePc plasma polymers deposited by RPAVD-$N_2$ at different polymerization degrees. (a) Sublimated FePc; (b) 30 W–3 cm; (c) 60 W; and (d) 240 W.

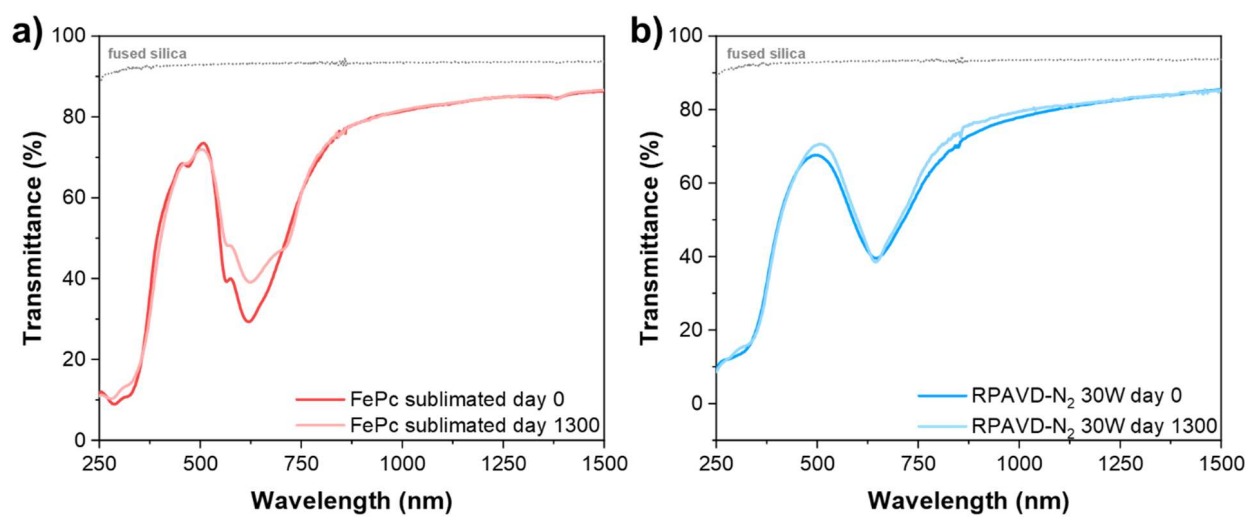

**Figure S4.** UV–Vis–NIR transmittance spectra of (a) sublimated FePc and (b) FePc30 W thin films, comparing as-deposited samples with those after air-exposure degradation for 1300 days.

The survey spectra and the atomic percentages of each element are shown in **Figures S5a and S5c**, respectively. Notable differences were observed in the relative amounts of carbon, nitrogen, and oxygen before and after $N_2$ plasma exposure. Specifically, the as-prepared CNFs contain C = 85.7%, N = 9.2%, O = 4.8%, and Na = 0.3%; whereas the plasma-treated CNFs exhibit C = 69.5%, N = 18.0%, O = 10.4%, and Na = 2.0%. These results confirm the incorporation of nitrogen-containing functional groups from the plasma into the fibrous matrix. The presence of sodium is most likely due to contamination introduced during handling, stabilisation or carbonisation.

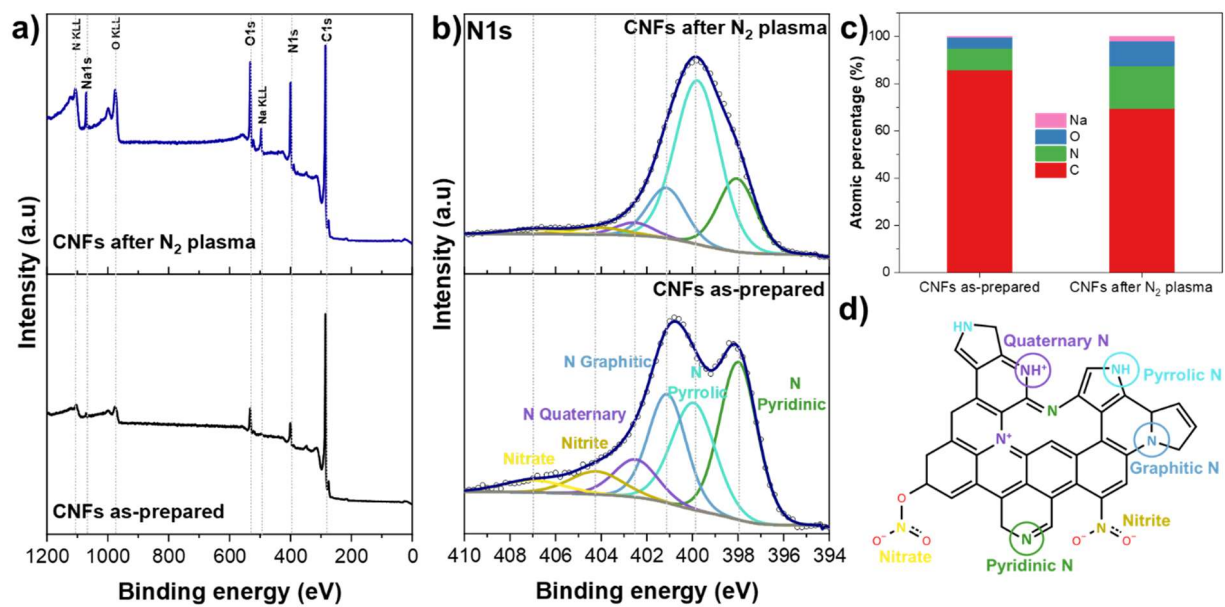

**Figure S5.** (a) XPS survey spectra of CNFs before and after $N_2$ plasma pretreatment. (b) High-resolution N1s spectra highlighting the evolution of nitrogen species upon plasma exposure. (c) Atomic percentages of C, N, O, and Na quantified from the XPS data. Both samples display the characteristic features of nitrogen-doped carbon materials, including pyridinic N (398.0 eV), pyrrolic N (399.9 eV), graphitic N (401.1 eV), and positively charged quaternary N (402.5 eV), as well as oxidized nitrogen species, namely $-NO_2$ (404.2 eV) and $-NO_3$ (407.0 eV).[1] After $N_2$ plasma treatment, the pyrrolic-N component increases markedly at the expense of the other species, indicating that the plasma induces a soft etching process that partially fragments the carbon rings. The highly reactive $N_2$ species then react with the resulting carbon radicals, promoting the incorporation of additional nitrogen and the formation of new pyrrolic-type sites. This nitrogen enrichment, particularly in pyrrolic-type sites, is expected to influence the coordination behaviour of FePc upon deposition. (d) Schematic representation of nitrogen functionalities and defect sites introduced into the CNF surface.

(1) Hellgren, N.; Haasch, R. T.; Schmidt, S.; Hultman, L.; Petrov, I. Interpretation of X-Ray Photoelectron Spectra of Carbon-Nitride Thin Films: New Insights from in Situ XPS. *Carbon* **2016**, *108*, 242–252. https://doi.org/10.1016/j.carbon.2016.07.017.

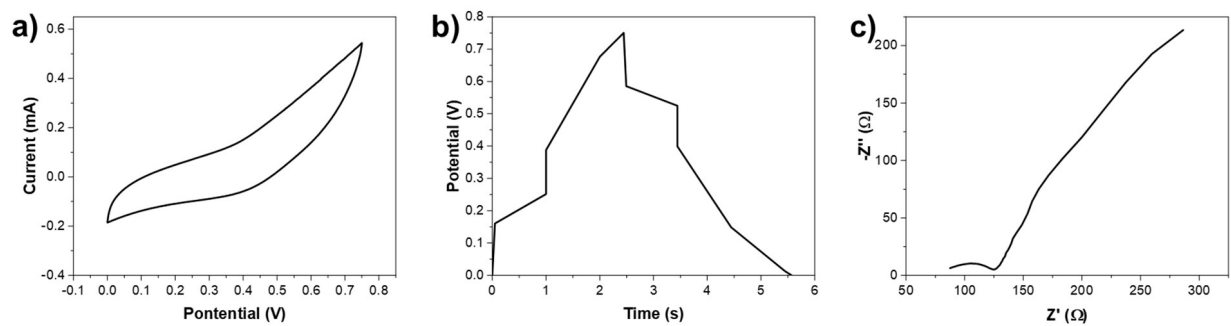

**Figure S6.** Electrochemical performance of $N_2$ plasma pre-treated CNTs. (a) Cyclic voltammetry curve recorded at a scan rate of $10\ mV \cdot s^{-1}$. (b) Galvanostatic charge-discharge profile measured at a current density of $0.25\ A \cdot g^{-1}$. (c) Electrochemical impedance spectroscopy (EIS) response.

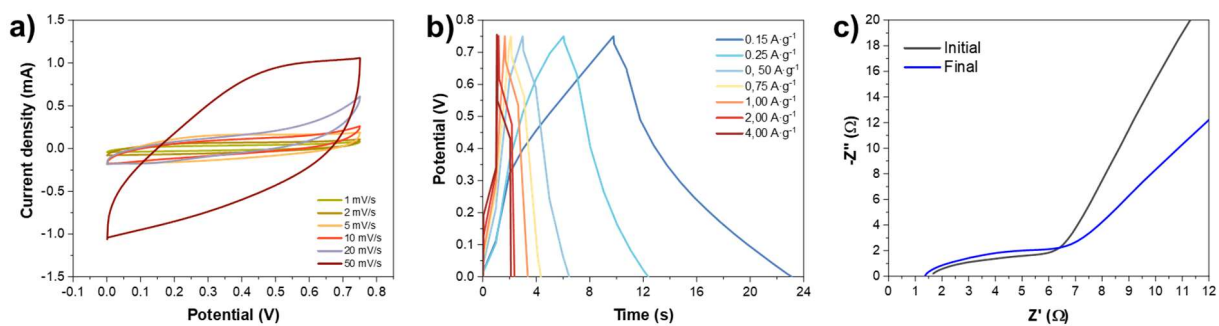

**Figure S7.** Electrochemical performance of FePc sublimated sample. (a) Cyclic voltammetry curve recorded at a scan rate of $10\ mV \cdot s^{-1}$. (b) Galvanostatic charge-discharge profile measured at a current density of $0.25\ A \cdot g^{-1}$. (c) Electrochemical impedance spectroscopy (EIS) response.

**Table S1.** Coefficients of the pseudocapacitive model $i(v)=k_1 \cdot v + k_2 \cdot v^{1/2}$ at selected voltages.

| Potential | $k_1$ ($mA \cdot mV^{-1} \cdot s^{-2}$) | Error $k_1$ | $k_2$ ($mA \cdot mV^{-1} \cdot s^{-1/2}$) | Error $k_2$ | $R^2$ |
|---|---|---|---|---|---|
| **0.00 V** | -0.069850 | ±0.010320 | -0.161090 | ±0.060580 | 0.99072 |
| **0.40 V (forward)** | 0.079900 | ±0.008600 | 0.152616 | ±0.050600 | 0.99480 |
| **0.75 V** | 0.077922 | ±0.006861 | 0.213459 | ±0.040260 | 0.99697 |
| **0.40 V (reverse)** | -0.075000 | ±0.009700 | -0.139800 | ±0.056700 | 0.99250 |

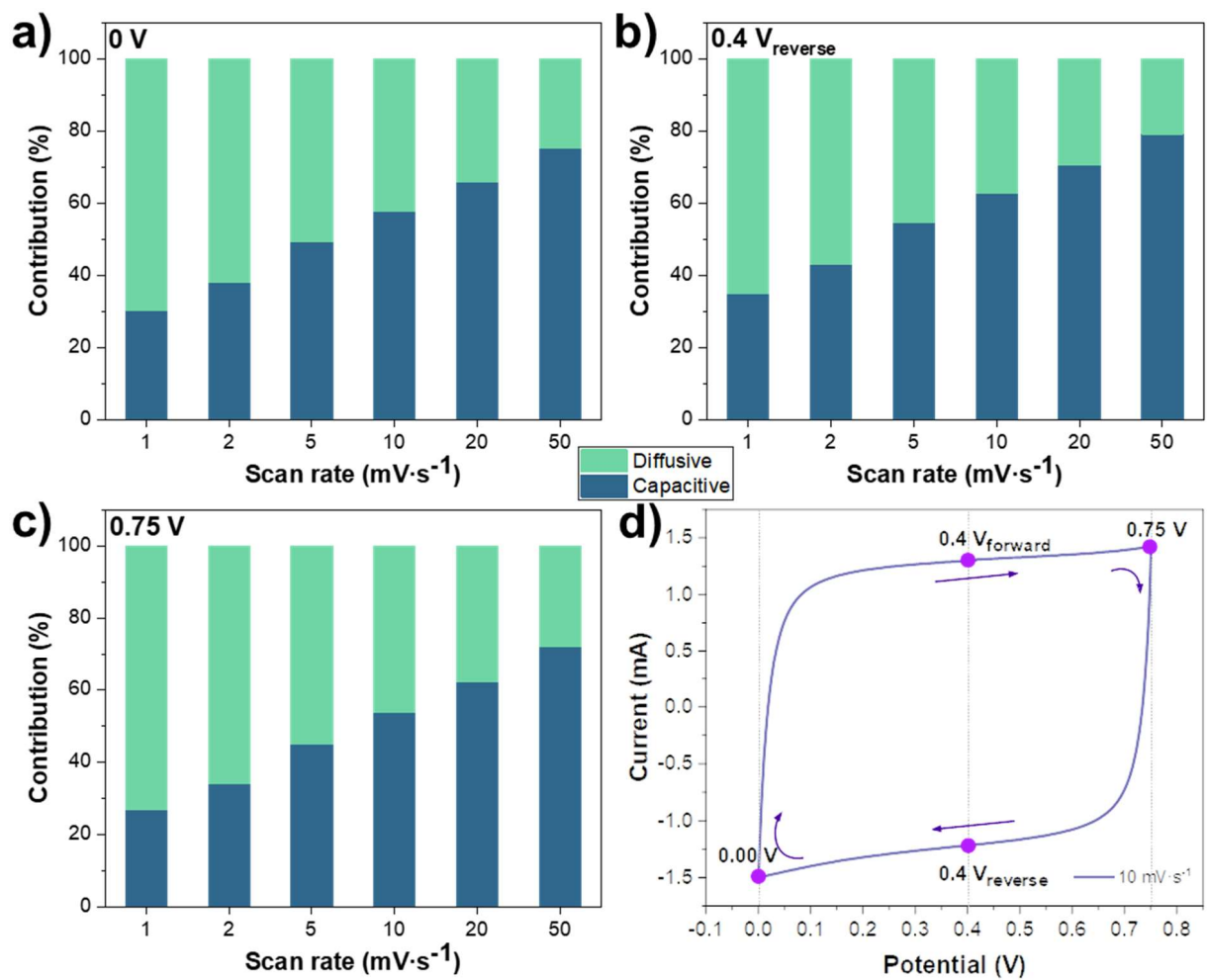

**Figure S8.** Deconvolution of capacitive and diffusive contributions at selected potentials for FePc30W@CNFs electrodes. Bar graphs showing the relative contributions of capacitive (blue) and diffusive (green) currents at scan rates from 1 to 50 mV·s⁻¹, extracted at: (a) 0.00 V, (b) 0.40 V (reverse scan) and (c) 0.75 V. Cyclic voltammogram highlighting the selected potentials (magenta dots) used for the deconvolution analysis.

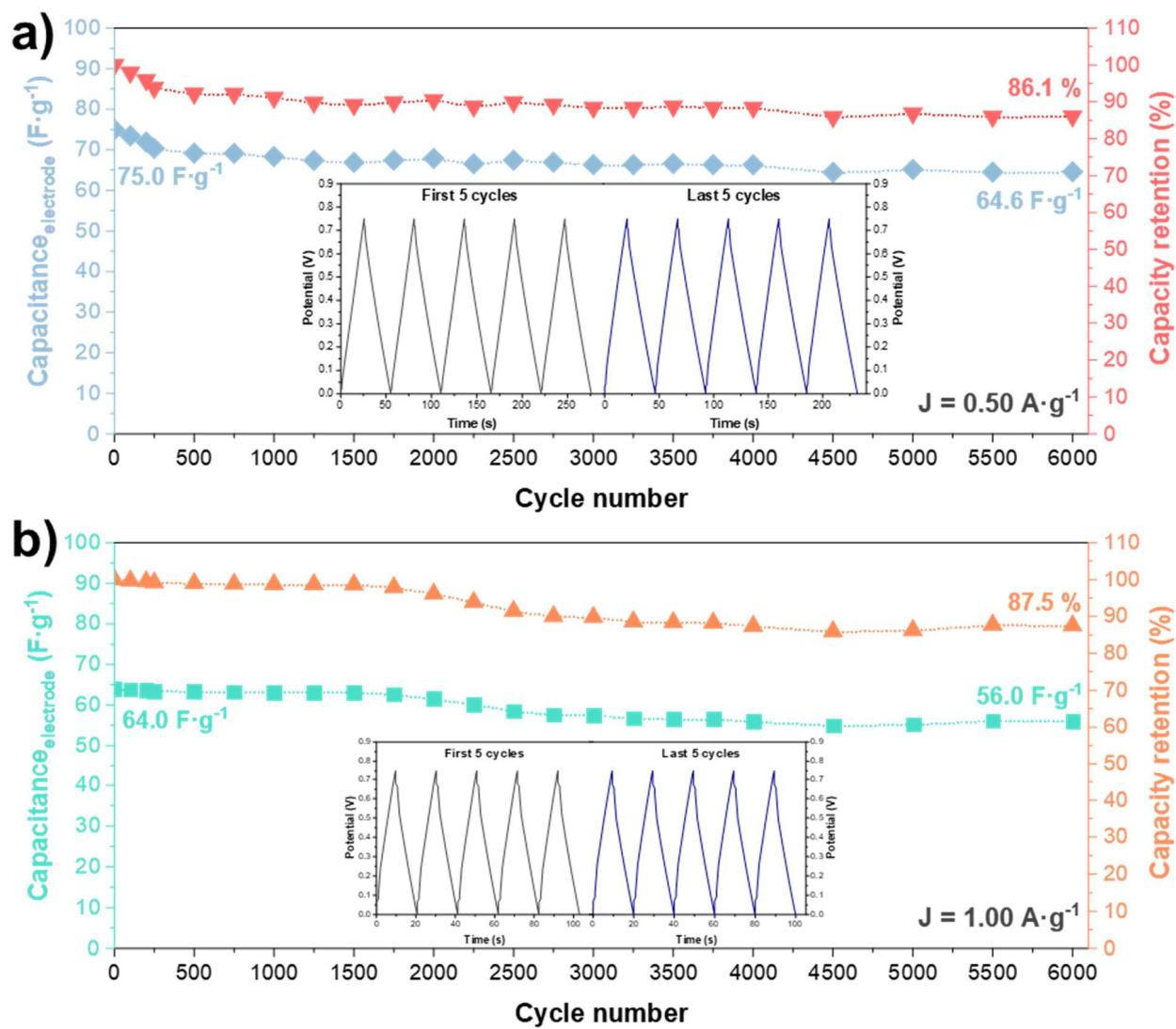

**Figure S9.** Cycling stability and capacitance retention of FePc30W@CNFs at (a) 0.50 A·g⁻¹ and (b) 1.00 A·g⁻¹. Each inset shows the first and last five cycles during the stability test.

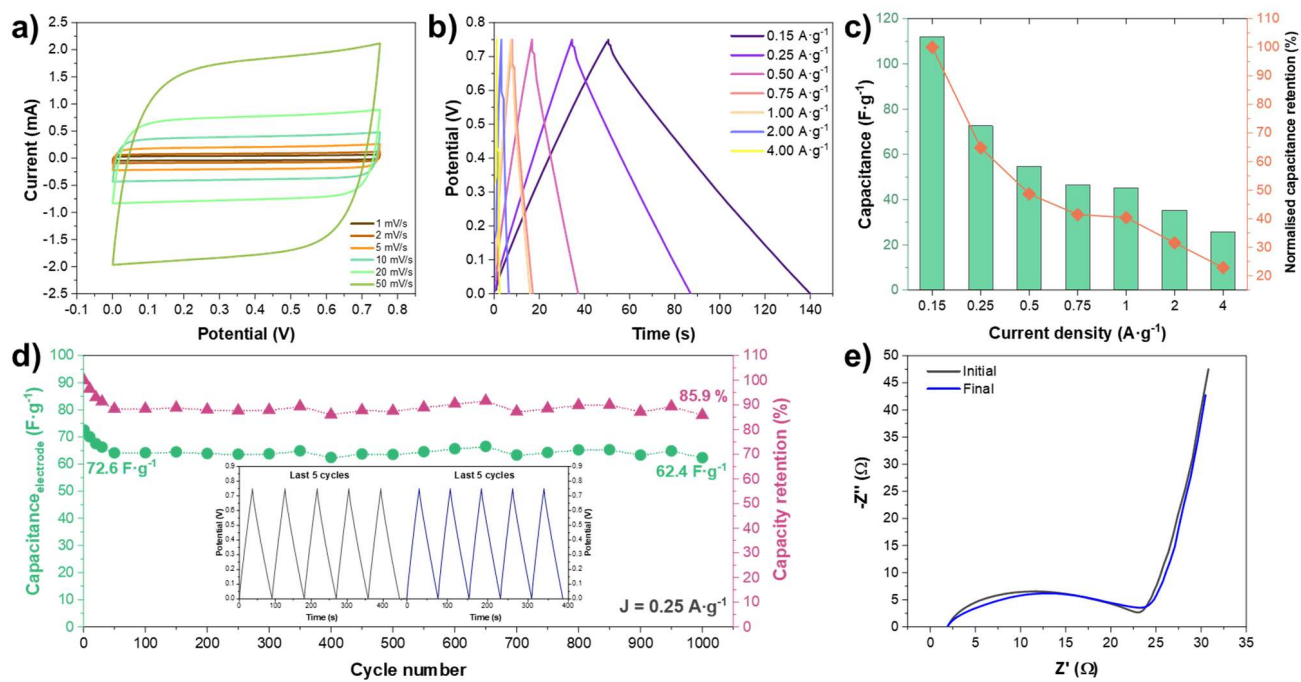

**Figure S10.** Electrochemical performance of FePc60W@CNFs sample. (a) CV curves recorded at scan rates ranging from 1 to 50 mV·s⁻¹. (b) GCD curves recorded at various current densities. (c) Specific capacitance ($C_{electrode}$) as a function of current density, along with the normalised capacitance relative to the lowest current density (0.15 A·g⁻¹). (d) Cycling stability and capacitance retention at 0.25 A·g⁻¹. The inset shows the first and last five cycles during the stability test. (e) Nyquist plots recorded before and after cycling.

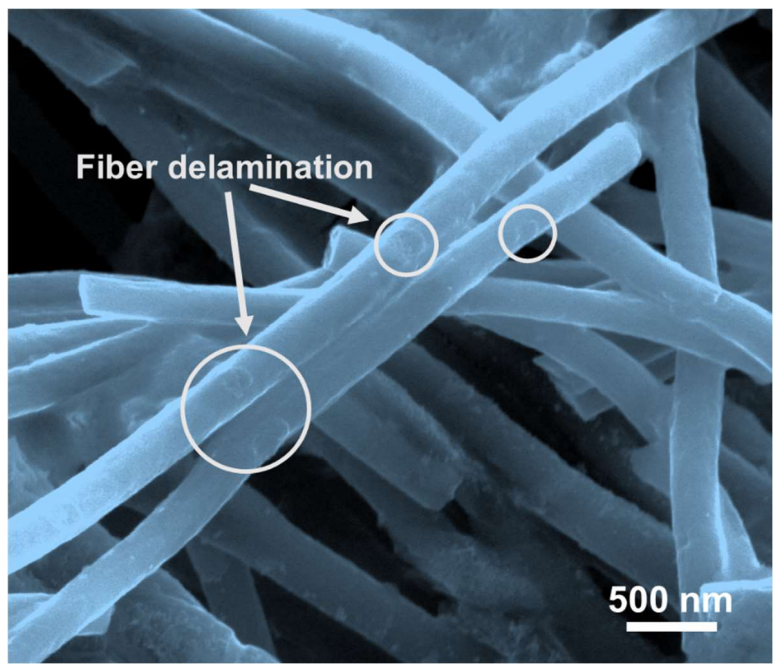

**Figure S11.** SEM micrographs of post-treated FePc30W@CNFs sample after 6000 charge-discharge cycles. Regions of partial delamination of the plasma-polymer coating have been labelled, attributed to electrode-electrolyte separation during cycling, which leads to the exposure of bare fiber areas.

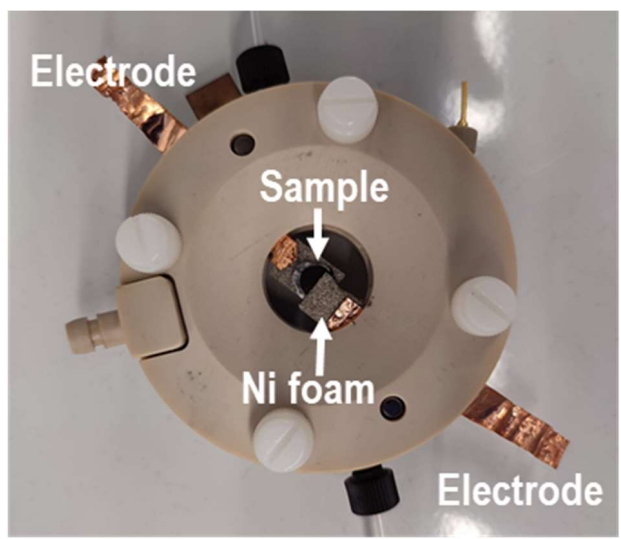

**Figure S12**. Top-view of the experimental set-up of *in situ* Raman measurements.

**Table S2.** Comparison of the result of this study with previous similar works on supercapacitor based in metal-phathalocyanines.

| Electrode type | Method | Specific Capacitance | Energy Density | Power Density | Cycling stability | | Electrolyte | Ref. |
|---|---|---|---|---|---|---|---|---|
| | | | | | No. of cycles | Capacitance retention (%) | | |
| FePc@C-HTC | Hydrothermal carbonisation + pyrolisis | 48 F·g$^{-1}$ at 0.2 A·g$^{-1}$ | 1.7 Wh·kg$^{-1}$ | 140.3 W·kg$^{-1}$ | 10000 at 5 A·g$^{-1}$ | 99 | 2M H$_2$SO$_4$ | [15] |
| FePc@GO | Self-assembly via π-π interactions | 235.5 F·g$^{-1}$ at 1 A·g$^{-1}$ | 8.2 Wh·kg$^{-1}$ | 531.9 W·kg$^{-1}$ | 60000 at 5 A·g$^{-1}$ | 100 | 6M KOH | [16] |
| FePc@NC-1000 | MOF synthesis + CVD carbonisation + FePc assembly | 0.03 F·cm$^{-2}$ at 5.96 A·cm$^{-2}$ | n/a | 0.12 W·cm$^{-2}$ | 10 h duration | 98.4 | Alcaline solution | [66] |
| FePc@MWNCTs | one step in-situ precipitation | 65.3 F·g$^{-1}$ at 0.25 A·g$^{-1}$ | 29.7 Wh·kg$^{-1}$ | 125.0 W·kg$^{-1}$ | 30000 at 8 A·g$^{-1}$ | 111.3 | 1M H$_2$SO$_4$ | [65] |
| CoPc@CNTs | acid-treated CNTs + CoPc ultrasonication | 1112 F·g$^{-1}$ at 1 A·g$^{-1}$ | 55.6 Wh·kg$^{-1}$ | 122.0 W·kg$^{-1}$ | 1000 at 1 A·g$^{-1}$ | 95 | 6M KOH | [14] |
| nCuPc+TiN | Hydrotermal | 29.7 F·g$^{-1}$ at 0.25 A·g$^{-1}$ | 2.4 Wh·kg$^{-1}$ | 388 W·kg$^{-1}$ | 30000 at 0.5 A·g$^{-1}$ | 93.5 | 1M Na$_2$SO$_4$ | [67] |
| β-ZnPc molecular crystal | Ring-closure reaction+annealing at 400 ºC | 49.1 F·g$^{-1}$ at 0.10 A·g$^{-1}$ | 86.2 Wh·kg$^{-1}$ | 220 W·kg$^{-1}$ | 100000 at 2 A·g$^{-1}$ | 73.4 | PVA/Zn(CF$_3$SO$_3$)$_2$ gel | [68] |
| NiPc@MWCNT | MWCNTs + NiPc ultrasonication | 0.19 F·cm$^{-2}$ at 1 mA·cm$^{-2}$ | 0.05 Wh·cm$^{-2}$ | 0.7 W·cm$^{-2}$ | 5000 at 10 mA·cm$^{-2}$ | 60 | PBS/KFC | [11] |
| CoPc@MWCNT | MWCNTs + CoPc ultrasonication | 0.51 F·cm$^{-2}$ at 1 mA·cm$^{-2}$ | 0.14 Wh·cm$^{-2}$ | 0.7 W·cm$^{-2}$ | 5000 at 10 mA·cm$^{-2}$ | 92 | PBS/KFC | [11] |
| NiPc NF-rGO | NiPc nanofibre synthesis + rGO hydrothermal assembly | 223.3 F·g$^{-1}$ at 1 A·g$^{-1}$ | 15.2 Wh·kg$^{-1}$ | 349.9 W·kg$^{-1}$ | 1000 at 1 A·g$^{-1}$ | n/a | 1M H$_2$SO$_4$ | [69] |
| NiPc@C-HTC | Hydrothermal carbonisation + pyrolisis | 48 F·g$^{-1}$ at 0.2 A·g$^{-1}$ | ≈1.6 Wh·kg$^{-1}$ | ≈120 W·kg$^{-1}$ | 10000 at 5 A·g$^{-1}$ | 98 | 2M H$_2$SO$_4$ | [15] |
| **Plasma polymer of FePc@electrospun CNFs** | **Remote plasma N$_2$ assisted vapour deposition** | **80.9 F·g$^{-1}$ at 0.25 A·g$^{-1}$ and 0.79 F·cm$^{-2}$ at 4.87 mA·cm$^{-2}$** | **6.3 Wh·kg$^{-1}$ and 6.2·10$^{-5}$ Wh·cm$^{-2}$** | **375 W·kg$^{-1}$ and 3.6·10$^{-4}$ W·cm$^{-2}$** | **6000 at 1 A·g$^{-1}$ and 6000 at 19.4 mA·cm$^{-2}$** | **87.5** | **6M KOH** | **This work** |